\documentclass[12pt,a4paper]{article}
\usepackage{graphicx}
\usepackage[T1]{fontenc}
\usepackage[utf8]{inputenc}
\usepackage{textcomp}
\usepackage[sc,osf]{mathpazo}
\usepackage{a4wide}  
\usepackage{latexsym,amsthm,amsfonts,amsmath,mathrsfs,amssymb}
\usepackage{accents}
\usepackage[unicode,implicit]{hyperref}
\hypersetup{%
  pdftitle    = {$\alpha'$ corrections to self-dual gravitational instantons}
  pdfkeywords = {gravitational instantons, D-instantons, string effective action, alpha
    prime corrections},
  pdfauthor   = {Jos\'e Luis V.~Cerdeira, Crist\'obal Corral and Tom\'as Ort\'{\i}n},
  plainpages  = true,
  colorlinks  = true,
  citecolor   = blue,
  urlcolor    = red,
  linkcolor   = blue
}
\newcommand{\hepth}[1]{{\tt
\href{http://www.arXiv.org/abs/hep-th/#1}{hep-th/#1}}}
\newcommand{\grqc}[1]{{\tt
\href{http://www.arXiv.org/abs/gr-qc/#1}{gr-qc/#1}}}

\newcommand{\arxiv}[1]{{\tt arXiv:\href{http://www.arXiv.org/abs/#1}{#1}}}
\newcommand{\doi}[1]{{\tt DOI:\href{http://dx.doi.org/#1}{#1}}}

\allowdisplaybreaks

\makeatletter
\@addtoreset{equation}{section}
\makeatother

\begin{document}

\begin{flushright}
\small
IFT-UAM/CSIC-26-049\\
May 14\textsuperscript{th}, 2026\\
\normalsize
\end{flushright}

\begin{center}

  {\Large {\bf $\alpha'$ corrections to self-dual gravitational instantons
}}
 
\vspace{1cm}

\renewcommand{\thefootnote}{\alph{footnote}}

 Jos\'e Luis V.~Cerdeira,$^{1,}$\footnote{Email: {\tt jose.verez-fraguela [at]
     estudiante.uam.es}}
 Crist\'obal Corral,$^{2,}$\footnote{Email: {\tt cristobal.corral [at]
     uai.cl}}
 and Tom\'as Ort\'{\i}n$^{1,}$\footnote{Email: {\tt Tomas.Ortin [at] csic.es}}

\setcounter{footnote}{0}
\renewcommand{\thefootnote}{\arabic{footnote}}
\vspace{.5cm}

${}^{1}${\it\small Instituto de F\'{\i}sica Te\'orica UAM/CSIC\\
  C/ Nicol\'as Cabrera, 13--15, C.U.~Cantoblanco, E-28049 Madrid, Spain}

  \vspace{0.2cm}

${}^{2}${\it\small Departamento de Ciencias, Facultad de Artes Liberales,
  Universidad Adolfo Ib\'a\~nez,\\ Avenida Padre Hurtado 750, 2562340, Vi\~na
  del Mar, Chile}

\vspace{1cm}

{\bf Abstract}
\end{center}
\begin{quotation} {\small We study the $\alpha'$ corrections to self-dual
    gravitational instantons in the context of the four-dimensional
    Cano--Ruip\'erez action, which can be obtained by the compactification of
    the Bergshoeff--de Roo heterotic string effective action on
    $\mathbb{T}^{6}$ followed by a truncation and a field redefinition. We
    show that the metric of spaces of self-dual curvature does not receive any
    corrections, but their (initially trivial) dilaton and axion fields do,
    owing to their couplings to Gauss--Bonnet and Pontrjagin densities. We
    find the generic form of the corrections of the dilaton and axion fields
    for the Gibbons--Hawking multi-instanton solutions and their explicit form
    for the particular cases of the Euclidean Taub--NUT and Eguchi--Hanson
    spaces. We construct the boundary terms required to define a well-posed
    Dirichlet variational principle in the Euclidean Cano--Ruip\'erez theory,
    including the contributions associated with the Gauss--Bonnet and
    Pontrjagin terms. The boundary terms are normalized for
    asymptotically-locally-Euclidean solutions, and we evaluate with them the
    Euclidean action of the $\alpha'$-corrected Eguchi--Hanson instanton
    showing that the total action receives no corrections to first order in
    $\alpha'$. We also show that, at zeroth order in $\alpha'$, one can
    construct Euclidean solutions similar to the string theory D-instanton
    with non-trivial dilaton and axion on the background of a self-dual purely
    gravitational instanton which remains unmodified. We also compute the
    $\alpha'$ corrections to these solutions.
  }
\end{quotation}

\newpage
\pagestyle{plain}



\section{Introduction}

Higher-curvature corrections are unavoidable if general relativity is
treated as an effective field theory. The string low-energy
effective field theory is a double series in the string coupling
constant $g_{s}$ that counts string loops, and in the Regge slope
parameter $\alpha'=\ell_{s}^{2}$ ($\ell_{s}$ is the string length),
that counts $\sigma$-model loops and finite-size string effects. At
lowest order, this effective action contains the Einstein-Hilbert term
\cite{Scherk:1974ca} and higher-order terms in $\alpha'$ (the
only ones that have been computed) contain higher derivative and
curvature terms
\cite{Zwiebach:1985uq,Callan:1985ia,Gross:1986iv,Gross:1986mw,Metsaev:1987zx,Hull:1987pc,Bergshoeff:1988nn,Bergshoeff:1989de}.

Using arguments of local supersymmetry, the heterotic superstring
effective action was computed in Ref.~\cite{Bergshoeff:1989de} to
cubic order in $\alpha'$. After compactification on $\mathbb{T}^{6}$,
consistent truncations and field redefinitions, a simple 4-dimensional
version of the first-order in $\alpha'$ Bergshoeff--de Roo action was
found in Ref.~\cite{Cano:2021rey}, describing the dynamics of the
metric, dilaton, and axion fields. The latter is the 4-dimensional dual
of the Kalb--Ramond field. As previous works indicated
\cite{Boulware:1986dr,Kanti:1995vq,Torii:1996yi,Alexeev:1996vs}, in
the action found in Ref.~\cite{Cano:2021rey}, which we will henceforth
refer to as the \textit{Cano--Ruip\'erez action}, the dilaton couples
to the Gauss--Bonnet density as in the popular dilaton-Gauss--Bonnet
models. However, it also contains the coupling of the axion to the
Pontrjagin density. This often-ignored coupling resembles
Chern--Simons-type modifications to general
relativity~\cite{Campbell:1990fu,Jackiw:2003pm,Alexander:2009tp}, for
which different black hole solutions are
known~\cite{Grumiller:2007rv,Konno:2009kg,Ahmedov:2010fz,Brihaye:2016lsx,Cisterna:2018jsx,Nashed:2023qjm}
and cannot be consistently eliminated from the string effective
action. In other words, contrary to popular belief, the only solutions
of the Einstein-dilaton Gauss--Bonnet models, which are also solutions
of the string theory effective action are those for which the
Pontrjagin density vanishes identically. In 4 dimensions, to first
order in $\alpha'$, one needs to work with the full Cano--Ruip\'erez
action.

Once the corrections to the equations of motion are known, one can
study how they modify the solutions to the uncorrected (lowest-order
in $\alpha'$) equations of motion. All the solutions of the
4-dimensional vacuum Einstein equations are also solutions of the
lowest-order Cano--Ruip\'erez string effective action with constant
dilaton and axion fields, and finding the $\alpha'$ corrections to the
metric and to these constant dilaton and axion fields is an
interesting problem because the deviations from the original solutions
can be understood as stringy effects. The first-order $\alpha'$
corrections to the Kerr black hole were studied semi-analytically in
Ref.~\cite{Cano:2021rey}.\footnote{Higher-curvature corrections to
  stationary configurations have been studied mainly numerically, as
  the highly nonlinear nature of the field equations complicates their
  analytic derivation. However, there are a few examples where
  analytic stationary solutions can be obtained: rotating black holes
  in Einstein--Gauss--Bonnet gravity at the Chern--Simons
  point~\cite{Anabalon:2024abz,Tapia:2025vtn}, boosted rotating black
  branes with arbitrary Gauss--Bonnet
  coupling~\cite{Dehghani:2002wn,Dehghani:2003ea,Dehghani:2006cu,Hendi:2010zza}.}
Their extremal limit was analyzed in Ref.~\cite{Cano:2023dyg} by
considering a potential that stabilizes the dilaton field. In the
absence of axions, the corrections to slowly rotating and accelerating
black holes were studied in Ref.~\cite{Agurto-Sepulveda:2022vvf}, and
the shadows were later analyzed in
Ref.~\cite{Agurto-Sepulveda:2024iwu}.  The thermodynamics of the
$\alpha'$-corrected black-hole solutions was studied in
Ref.~\cite{Ortin:2024emt}, where it was shown that, both in the Smarr
formula and in the first law of black-hole mechanics, $\alpha'$ plays
the role of a thermodynamic variable, as it had already been noticed
in a slightly different context in Ref.~\cite{Zatti:2023oiq} in
agreement with the general arguments concerning dimensionful constants
in Refs.~\cite{Ortin:2021ade,Meessen:2022hcg}. Further work on the
$\alpha'$ corrections of charged black-hole solutions of the heterotic
superstring effective action, which cannot be studied using the
Cano--Ruip\'erez action, can be found in
Refs.~\cite{Cano:2018qev,Chimento:2018kop,Cano:2018brq,Cano:2019ycn,Cano:2021nzo,Ortin:2021win,Cano:2022tmn,Hu:2025aji}.

Gravitational instantons (globally regular solutions of the Euclidean
equations of motion with finite action) are another class of solutions
whose higher-order corrections are of great interest. Although
gravitational instanton solutions have been obtained and studied in
the context of Einstein--Gauss--Bonnet and Lovelock theories in
Refs.~\cite{Dehghani:2005zm,Dehghani:2006dh,Dehghani:2006aa,Hendi:2008wq,Corral:2019leh,Corral:2022udb,Corral:2025yvr},
the stringy $\alpha'$ corrections to the gravitational instantons of
general relativity have not yet been studied. Doing that in the
context of the Euclidean Cano-Ruip\'erez action is the main goal of
this paper.

Self-dual gravitational instantons of general relativity have been
exhaustively studied because they are easier to find, since self-duality and
the Bianchi identity automatically imply that the equations of motion
are satisfied. They are hyper-K\"ahler spaces, \textit{i.e.}~spaces of
special holonomy (self-duality implies that it is just
$\text{SU}(2)\in\text{SO}(4)$), which is a property which is related to unbroken
supersymmetry, and they saturate a
Bogomol'nyi--Prasad--Sommerfield-type bound that renders the Euclidean
action proportional to a topological invariant of the Pontrjagin
class.

The Gibbons--Hawking family of solutions, which can be constructed in
terms of a single harmonic function in $\mathbb{E}^{3}$,
\cite{Gibbons:1979zt,Gibbons:1987sp} contains all the self-dual
metrics admitting a triholomorphic isometry, many of which are
(multi-) gravitational instantons.\footnote{That is: an isometry that
  preserves the triholomorphic structure that characterizes a
  hyperK\"ahler space.} In particular, it contains the Euclidean
self-dual Taub--NUT
metric~\cite{Taub:1950ez,Newman:1963yy,Hawking:1976jb,Eguchi:1977iu},
the multi-Taub--NUT metric, and the Eguchi--Hanson
instanton~\cite{Eguchi:1978xp,Eguchi:1978gw} as particular cases (see
also~\cite{Eguchi:1980jx}) that arise from particular choices of
harmonic function. These solutions resemble 't Hooft's Yang--Mills
multi-instanton solutions \cite{tHooft:1976rip,tHooft:1976snw}.

In this paper, we will focus on the $\alpha'$ corrections to self-dual
gravitational instantons, in particular, Gibbons--Hawking spaces. As
solutions of the zeroth-order Euclidean Cano--Ruip\'erez action, they
have to be supplemented by constant axion and dilaton fields. However,
we will show that these trivial embeddings of self-dual instantons in
string theory can be generalized because it is possible to have a
non-constant combination of the axion and dilaton that solves the
equations of motion with exactly the same metric. The non-trivial
scalar combination is given by the inverse of a function which is
harmonic in the background of the gravitational instanton. These
solutions are nothing but a 4-dimensional generalization of the
10-dimensional D-instanton of Ref.~\cite{Gibbons:1995vg}, whose metric
is $\mathbb{E}^{10}$ in which $\mathbb{E}^{4}$ is being replaced by a
self-dual gravitational instanton.\footnote{In the D-instanton solution of
  Ref.~\cite{Gibbons:1995vg}, the axion is the Ramond--Ramond 0-form while in
  this setting it is the dual of the Neveu--Schwarz-Neveu--Schwarz
  Kalb--Ramond 2-form in 4 dimensions.} 

The first-order in $\alpha'$ corrections of these generalized D-instantons can
be computed for any choice of harmonic function and, as particularly
interesting examples.  We will determine the corrections when the metric is an
arbitrary Gibbons-Hawking space and for the particular cases of the Taub--NUT
and Eguchi--Hanson solutions for some particular choices of harmonic
functions.

Since the main properties of instantons are the finiteness of their
actions, we will complete the Euclidean Cano--Ruip\'erez action with boundary
terms such that the Dirichlet variational problem is well posed and, with
Dirichlet boundary conditions for the variations of the fields, the variation
of the action vanishes identically for solutions of the equations of
motion. Furthermore, the boundary terms will be normalized such that the value
of the action for the chosen vacuum will also vanish identically. This
completion of the Euclidean action is only adequate for asymptotically locally
Euclidean solutions and, therefore, we will only compute it for the
$\alpha'$-corrected Eguchi--Hanson studied before, finding that, if an inner
boundary is taken into account, the $\alpha'$ corrections to the action vanish
identically.

This paper is organized as follows: In Section~\ref{sec-thetheory}, we present
the Euclidean version of the Cano--Ruip\'erez action and its field
equations. Then, in Section~\ref{sec-selfdual}, we focus on self-dual spaces
and obtain a reduced system for the dilaton and axion fields. In
Section~\ref{sec-GHspaces}, we study $\alpha'$ corrections to the
Gibbons--Hawking spaces, alongside their particular limits to the Euclidean
Taub--NUT and Eguchi--Hanson spaces in Sections~\ref{sec-ETN}
and~\ref{sec-EguchiHanson}, respectively. In
Section~\ref{sec-euclideanaction}, we obtain the boundary terms for the
Cano--Ruip\'erez action and obtain the contribution of the Eguchi--Hanson
instanton. Finally, in Section~\ref{sec-discussion}, we present our
conclusions and closing remarks.

\section{The Euclidean Cano--Ruip\'erez string effective action}
\label{sec-thetheory}

In Ref.~\cite{Cano:2021rey}, starting with the Bergshoeff-de Roo first order
in $\alpha'$, 10-dimensional Heterotic String effective action
\cite{Bergshoeff:1989de} and its compactification on $\mathbb{T}^{6}$, Cano and
Ruip\'erez found the following expression for the 4-dimensional effective
action for the Vierbein $e^{a}=e^{a}{}_{\mu}dx^{\mu}$, dilaton $\phi$ and
axion\footnote{The 4-dimensional axion field is the dual of the Kalb--Ramond
  2-form.} $\chi$ fields \cite{Ortin:2024emt}

\begin{equation}
  \label{eq:CRaction}
\begin{aligned}
  S_{\rm CR}[e^{a},\phi,\chi]
  & =
    \frac{1}{16\pi G_{N}} \int \left\{ -\star(e^{a}\wedge e^{b}) \wedge R_{ab}
    +\tfrac{1}{2}\left(d\phi \wedge \star d\phi
    +e^{2\phi} d\chi \wedge \star d\chi\right)
    \right.
  \\
    & \\
  & \hspace{.5cm}
    \left.
    -\frac{\alpha'}{4}\left(e^{-\phi} \tilde{R}^{ab}\wedge R_{ab}
    -\chi R^{a}{}_{b}\wedge R^{b}{}_{a}\right) \right\}\,,
\end{aligned}
\end{equation}

\noindent
where $G_N$ is the Newton's constant, $\star$ is the Hodge dual, $\wedge$ is the wedge product of differential forms, and the dual curvature two-form is defined as 

\begin{equation}
  \tilde{R}^{ab}
  \equiv
  \tfrac{1}{2}\varepsilon^{abcd}R_{cd}\,.
\end{equation}

The expression in Eq.~\eqref{eq:CRaction} is written in the conventions of
Refs.~\cite{Ortin:2015hya,Gomez-Fayren:2023qly} and in particular, using a
Lorentzian mostly-minus metric. Wick-rotating the time coordinate and taking
into account the pseudoscalar nature of the axion field the Cano--Ruip\'erez
action becomes\footnote{If the Kalb-Ramond field is a 2-form, its dual, the
  axion field, must be a pseudoscalar and must be replaced by $i\chi_{\rm E}$
  under a Wick rotation.}

\begin{equation}
  \label{eq:ECRaction}
\begin{aligned}
  S_{\rm CR}[e^{a},\phi,\chi_{\rm E}]
  & =
  iS_{\rm ECR}[e^{a},\phi,\chi_{\rm E}]
  \\
  & \\
  S_{\rm ECR}[e^{a},\phi,\chi_{\rm E}]
  & =
    \frac{1}{16\pi G_{N}} \int \left\{ -\star(e^{a}\wedge e^{b}) \wedge R_{ab}
    +\tfrac{1}{2}\left(d\phi \wedge \star d\phi
    -e^{2\phi} d\chi_{\rm E} \wedge \star d\chi_{\rm E}\right)
    \right.
  \\
    & \\
  & \hspace{.5cm}
    \left.
    -\frac{\alpha'}{4}\left(e^{-\phi} \tilde{R}^{ab}\wedge R_{ab}
    -\chi_{\rm E} R^{a}{}_{b}\wedge R^{b}{}_{a}\right) \right\}\,,
\end{aligned}
\end{equation}

\noindent
where, now, the metric used to raise and lower Lorentz indices is
$-\delta_{ab}$. Replacing that metric by the standard Euclidean metric
$\delta_{ab}$ the Euclidean action takes the form\footnote{The spin connection
  $\omega^{a}{}_{b}$ and the Lorentz curvature $R^{a}{}_{b}$ are left
  invariant by this change. Thus, $\omega^{ab}$ and $R^{ab}$ change their
  signs.}

\begin{equation}
  \label{eq:ECRaction2}
\begin{aligned}
  S_{\rm ECR}[e^{a},\phi,\chi_{\rm E}]
  & =
    \frac{1}{16\pi G_{N}} \int \left\{ \star(e^{a}\wedge e^{b}) \wedge R_{ab}
    -\tfrac{1}{2}\left(d\phi \wedge \star d\phi
    -e^{2\phi} d\chi_{\rm E} \wedge \star d\chi_{\rm E}\right)
    \right.
  \\
    & \\
  & \hspace{.5cm}
    \left.
    -\frac{\alpha'}{4}\left(e^{-\phi} \tilde{R}^{ab}\wedge R_{ab}
    -\chi_{\rm E} R^{a}{}_{b}\wedge R^{b}{}_{a}\right) \right\}\,.
\end{aligned}
\end{equation}

\noindent
Here, $\tilde{R}^{ab}\wedge R_{ab}$ and $R^{a}{}_{b}\wedge R^{b}{}_{a}$ are
the Gauss--Bonnet and the Pontrjagin densities, respectively. If the manifold
has no boundaries, their integrals are related to topological invariants: the
former to the Euler characteristic, and the latter to the Chern-Pontrjagin
index. In the presence of boundaries, their Chern(-Simons) forms and nonlocal
terms, such as the $\eta$-invariant, must be included in order to obtain the
corresponding topological invariants (see~\cite{Eguchi:1980jx} for
details). Although they can be locally written as total derivatives, because
of their couplings to the dilaton and axion fields, they do contribute to the
field dynamics. This is exactly the class of $\alpha'$ corrections we are
interested in, which we study throughout this work.

Defining now the two real scalar fields

\begin{equation}
  \phi^{(\pm)}
  \equiv
  e^{-\phi} \pm \chi_{\rm E}\,,
\end{equation}

\noindent
it is not difficult to see that the action can be rewritten in the form

\begin{equation}
  \label{eq:ECRaction3}
\begin{aligned}
  S_{\rm ECR}[e^{a},\phi^{(+)},\phi^{(-)}]
  & =
    \frac{1}{16\pi G_{N}} \int \left\{ \star(e^{a}\wedge e^{b}) \wedge R_{ab} %
    -2 \frac{d\phi^{(+)} \wedge \star d\phi^{(-)}}{\left[\phi^{(+)}+\phi^{(-)}\right]^{2}}
    \right.
  \\
    & \\
  & \hspace{.5cm}
    \left.
    -\frac{\alpha'}{4}\left[\phi^{(+)}R^{(+)\, ab}\wedge R^{(+)}{}_{ab}
    -\phi^{(-)}R^{(-)\, ab}\wedge R^{(-)}{}_{ab}\right] \right\}\,,
\end{aligned}
\end{equation}

\noindent
where we have defined the self- and anti-self-dual parts of the Lorentz
curvature\footnote{Notice that these are not the curvatures of the torsionful
  spin connections of Ref.~\cite{Bergshoeff:1989de}.}

\begin{equation}
  R^{(\pm)\, ab}
  \equiv
  \tfrac{1}{2}\left( R^{ab}\pm \tilde{R}^{ab}\right)\,.
\end{equation}

We are going to use this form of the action because it is clearly well-suited
to work with self- and anti-self-dual instanton solutions. This action is
enough to derive the classical equations of motion, but an action leading to a
well-defined variational principle is necessary in the path-integral
formulation of the quantum theory, where the actual value of the action has to
be extremized by the solutions to the classical equations of motion. As it is
well known, in order to have a well-posed variational principle, we will have
to supplement this action by an appropriate boundary term. We will deal with
this problem in Section~\ref{sec-euclideanaction}.

Under a generic variation of the fields

\begin{equation}
  \label{eq:ECRaction3variation}
    \delta S_{\rm ECR}
     =
      \int
      \left\{
      \mathbf{E}_{a} \wedge \delta e^{a}
      +\mathbf{E}_{(+)} \, \delta \phi^{(+)}
      +\mathbf{E}_{(-)} \, \delta \phi^{(-)}
      +d\mathbf{\Theta}(\varphi,  \delta \varphi)
      \right\}\,,
\end{equation}

\noindent
where 

\begin{subequations}
  \begin{align}
    \mathbf{E}_{a}
    & =
      -\imath_{a}\star (e^{b}\wedge e^{c})\wedge R_{bc}
      \nonumber \\
    & \nonumber \\
    & \hspace{.5cm}
      -\frac{1}{\left[\phi^{(+)}+\phi^{(-)}\right]^{2}}
      \left\{
      \imath_{a}d\phi^{(+)}\star d\phi^{(-)}
      +d\phi^{(+)}\wedge \imath_{a}\star d\phi^{(-)}
      +(+\leftrightarrow -)
      \right\}
      \nonumber \\
    & \nonumber \\
    & \hspace{.5cm}
   -\mathcal{D}\Delta_{a}   \,,
    \\
    & \nonumber \\
    \mathbf{E}_{(\pm)}
    & =
      2\frac{d\star d\phi^{(\mp)}}{\left[\phi^{(+)}+\phi^{(-)}\right]^{2}}
      -4\frac{d\phi^{(\mp)}\wedge \star
      d\phi^{(\mp)}}{\left[\phi^{(+)}+\phi^{(-)}\right]^{3}}
      \mp\frac{\alpha'}{4}R^{(\pm)\, ab}\wedge R^{(\pm)}{}_{ab}\,,
  \end{align}
\end{subequations}

\noindent
are the (left-hand-sides of the) equations of motion and 

\begin{equation}
  \label{eq:presymplecticpotential}
  \begin{aligned}
    \mathbf{\Theta}(\varphi,  \delta \varphi)
    & =
    \left[\star (e^{a}\wedge e^{b}) +\mathbf{H}_{ab}\right]\wedge \delta \omega_{ab}
    -2 \frac{\star d\phi^{(+)}}{\left[\phi^{(+)}+\phi^{(-)}\right]^{2}}\delta \phi^{(-)}
    -2 \frac{\star d\phi^{(-)}}{\left[\phi^{(+)}+\phi^{(-)}\right]^{2}}\delta \phi^{(+)}
    \\
    & \\
    & \hspace{.5cm}
      +\Delta_{a}\wedge \delta e^{a}\,,
  \end{aligned}
\end{equation}

\noindent
is the pre-symplectic potential 3-form ($\varphi$ represents all the
independent fields of the theory). In the Einstein equation, $\mathbf{E}_{a}$,
and in the presymplectic potential, $\mathbf{\Theta}(\varphi,\delta\varphi)$,
the 2-form $\Delta_{a}$ is defined using the terms of first order in $\alpha'$
of the variation of the action with respect to the connection
$\mathbf{E}^{(1)}{}_{a}{}^{b}$ as \cite{Ortin:2024emt}

\begin{equation}
  \Delta_{a}
  \equiv
  2\imath_{b}\mathbf{E}^{(1)\, b}{}_{a}
  -\tfrac{1}{2}\imath_{b}\imath_{c}\mathbf{E}^{(1)\, bc}\wedge e^{d}\delta_{da}\,.
\end{equation}

Here this variation is explicitly given by

\begin{subequations}
  \begin{align}
    \mathbf{E}^{(1)}{}_{a}{}^{b}
    & =
      -\mathcal{D}\mathbf{H}_{a}{}^{b}\,,
    \\
    & \nonumber \\
  \mathbf{H}_{a}{}^{b}
  & \equiv
  -\frac{\alpha'}{2}\left[\phi^{(+)}R^{(+)}{}_{a}{}^{b}
    -\phi^{(-)}R^{(-)}{}_{a}{}^{b}\right]
    \\
    & \nonumber \\
    \label{eq:Hab}
    & =
  -\frac{\alpha'}{2}\left[e^{-\phi}\tilde{R}_{a}{}^{b}
    +\chi_{E}R_{a}{}^{b}\right]\,.
  \end{align}
\end{subequations}

\section{Self-dual instantons}
\label{sec-selfdual}

In this work, we are going to focus on solutions to the zeroth-order Euclidean
action (instantons) with self-dual curvature

\begin{equation}
  R^{(-)\, ab}
  =
  0\,,
  \hspace{1cm}
    R^{(+)\, ab}
  =
  R^{ab}\,.
\end{equation}

Self- and anti-self-dual curvature 2-forms lead to vanishing Ricci and
Einstein 1-forms,\footnote{The Ricci and Einstein 1-forms can be defined as
      \begin{align}
      R_{\rm ic}^{a}
       \equiv
        \imath_{b}R^{ab}\,, \qquad \mbox{and} \qquad G^{a}
       \equiv
        R_{\rm ic}^{a} -\tfrac{1}{2}R_{\rm ic} e^{a}\,,  
    \end{align}
    respectively, where $R_{\rm ic} \equiv \imath_{a}R_{\rm ic}^{a}$ is the
    Ricci scalar. Then, taking the interior product $\imath_{a}$ of an (anti-)
    self-dual curvature
\begin{equation}
\imath_{a}\left(R^{ab}=\pm\tfrac{1}{2}\varepsilon^{abcd}R_{cd}\right)\,,  
\end{equation}
and using the Bianchi identity $\imath_{[a}R_{cd]}=0$, we see that the Ricci
1-form vanishes.} which means that either $\phi^{(+)}$ or $\phi^{(-)}$ (or
both) must be constant at zeroth order in $\alpha'$ for the Einstein equations
to be satisfied. If we set one of the scalars to a constant, the remaining
zeroth-order scalar equations of motion simplify dramatically:

\begin{equation}
 \left.   \mathbf{E}_{(\pm)}\right|_{\phi^{(\pm)}=C}
    =
      -2d\star d\frac{1}{\left[C+\phi^{(\mp)}\right]}\,,
\end{equation}

\noindent
and we conclude that, given a metric with self- or anti-self-dual curvature,
we can construct zeroth-order solutions of this theory by adding to the metric
a constant scalar and a scalar which is the inverse of a harmonic function, up
to an additive constant, in that space. At zeroth order, the theory is
completely symmetric under the interchange $\phi^{(+)}\rightarrow \phi^{(-)}$
and the choice of constant and non-constant scalar is completely arbitrary.

As mentioned in the introduction, these solutions resemble the 10-dimensional
D-instanton solution of the type~IIB supergravity effective action found in
Ref.~\cite{Gibbons:1995vg} in which only the dilaton and axion field are
active and given by a harmonic function and the metric is flat
$\mathbb{E}^{10}$. The solutions we have just found are non trivial when the
metric is that of $\mathbb{E}^{4}$ but they can also be constructed of any
other metric with self-dual curvature. In this sense, they generalize the
4-dimensional version of the D-instanton, although the axion we are dealing
with belongs to the NS-NS sector and they cannot be properly called
D-instantons.

Things become more complicated at first order in $\alpha'$ because, in
principle, all the fields can get $\alpha'$ corrections. However, we notice
that $\mathbf{E}^{(1)}{}_{a}{}^{b}$ can be set to zero for a metric with
self-dual curvature (our particular choice) by choosing $\phi^{(+)}$ to be the
constant scalar at zeroth order in $\alpha'$. The energy-momentum tensor of
the scalars can also be set to zero if we assume that $\phi^{(+)}$ is still
constant at first order in $\alpha'$,
\textit{i.e.}~$\phi^{(+)} = B +\alpha' C+ \mathcal{O}(\alpha^{\prime\, 2})$.
Thus, the Einstein equation will still be solved by the original self-dual
metric for any $\phi^{(-)}$. It only remains to be seen if and how we can solve
the scalar equations.

The equations of the two scalars are, now,

\begin{subequations}
  \begin{align}
      2\frac{d\star d\phi^{(-)}}{\left[\phi^{(+)}+\phi^{(-)}\right]^{2}}
    -4\frac{d\phi^{(-)}\wedge \star d\phi^{(-)}}
    {\left[\phi^{(+)}+\phi^{(-)}\right]^{3}}
    -\frac{\alpha'}{4}R^{ab}\wedge R_{ab}
    & =
      0\,,
    \\
    & \nonumber \\
      2\frac{d\star d\phi^{(+)}}{\left[\phi^{(+)}+\phi^{(-)}\right]^{2}}
    -4\frac{d\phi^{(+)}\wedge \star d\phi^{(+)}}
    {\left[\phi^{(+)}+\phi^{(-)}\right]^{3}}
        & =
      0\,.
  \end{align}
\end{subequations}

\noindent
Our choice $\phi^{(+)} = B +\alpha' C+ \mathcal{O}(\alpha^{\prime\, 2})$
solves the second and brings the first to the form

\begin{equation}\label{eq:phiminus}
  d\star d\frac{1}{\left[ B +\alpha' C+\phi^{(-)}\right]}
  +\frac{\alpha'}{8}R^{ab}\wedge R_{ab}
  =
  0\,.
\end{equation}

This equation can be solved order by order in $\alpha'$. To do so, we assume
that $\phi^{(-)}$ admits a series expansion as

\begin{equation}
    \phi^{(-)} = \phi^{(-)}_0 + \alpha'\,\phi^{(-)}_1 + \mathcal{O}(\alpha'^2)\,.
\end{equation}

\noindent
To zeroth order in $\alpha'$, the solution of Eq.~\eqref{eq:phiminus} is given by 

\begin{equation}
    \phi^{(-)}_0 = -B + \frac{1}{\mathcal{H}}\,,
\end{equation}

\noindent
where $\mathcal{H}$ is an harmonic function chosen at that order.
%
%
Then, using the fact that there are no $\alpha'$ corrections implicit in the Hodge star operator, the equation for the first-order terms reduces to

\begin{equation}
  \label{eq:forfirstorderterms}
  d\star d \left[\mathcal{H}^{2}\left(C+\phi_{1}^{(-)}\right) \right]
  -\tfrac{1}{8}R^{ab}\wedge R_{ab}
  =
  0\,.
\end{equation}

In order to solve this equation for the first-order correction
$\phi_{1}^{(+)}$ we need to specify the self-dual instanton we have started
from. For a wide class of them, though, it is possible to find a formal
solution, as we are going to see next.

\subsection{Gibbons--Hawking spaces}
\label{sec-GHspaces}

Gibbons--Hawking (GH) spaces are four-dimensional hyper-K\"ahler manifolds
admitting a triholomorphic isometry, while having a self-dual curvature
2-form~\cite{Gibbons:1979zt,Gibbons:1987sp}.\footnote{Different conventions
  and choices (orientation, signs) lead to anti-self-dual curvatures. Our
  choices, identical to those of Ref.~\cite{Chimento:2018kop} for the sake of
  simplicity, lead to self-dual curvatures.} The metric of GH spaces can
always be written in the form

\begin{equation}
d\sigma^{2}= H^{-1}(d\eta+\chi)^{2}+Hdx^{x}dx^{x}\, ,
\,\,\,\,\,
\partial_{\underline{x}}H
=
\varepsilon_{xyz}\partial_{\underline{y}}\chi_{\underline{z}}\,.  
\end{equation}

\noindent
Here, $\eta=x^{\sharp}$ is the coordinate adapted to the isometry and the rest
of the coordinates are denoted by $x^{x}$ with $x,y,z=1,2,3$. The tangent
space indices $a,b,\cdots $ take the values $\sharp,1,2,3$. Underlined indices
are coordinate-base (``world'') indices. The integrability condition of the
equation that relates $H$ and $\chi_{\underline{z}}$ (a 1-form in Euclidean
3-dimensional space $\mathbb{E}^{3}$) implies that $H$ is a harmonic form in
$\mathbb{E}^{3}$. Allowing for point-like singularities, $H$ can be
generically written in the form

\begin{equation}
  \label{eq:H-GHspaces}
  H
  =
  \epsilon +\sum_{n}\frac{a_{n}}{|\vec{x}-\vec{x}_{n}|}\,,
\end{equation}

\noindent
where $\epsilon$ and all the $a_{n}$ are real constants,
$\vec{x}=(x^{1},x^{2},x^{3})$ and
$\vec{x}_{n}=(x_{n}^{1},x_{n}^{2},x_{n}^{3})$ is a constant 3-vector.

In the frame

\begin{equation}
\label{eq:simplestframe}
\begin{array}{rclrcl}
v^{\sharp} & = & H^{-\frac{1}{2}}[d\eta +\chi_{\underline{x}}dx^{x}]\, ,
\hspace{1cm}
& 
v_{\sharp} & = & H^{\frac{1}{2}}\partial_{\eta}\equiv \partial_{\sharp}\, ,\\
& & & & & \\
v^{x} & = & H^{\frac{1}{2}}dx^{x}\, ,
& 
v_{x} & = & H^{-\frac{1}{2}}[\partial_{\underline{x}} 
-\chi_{\underline{x}}\partial_{\eta} ]=\partial_{x}\, ,\\
\end{array}
\end{equation}

\noindent
and using the matrices 

\begin{equation}
  (\mathbb{N}^{\pm}_{ab})_{cd} \equiv
  \eta_{ae}\eta_{bf}
  \left(\delta_{cd}{}^{ef}\pm \tfrac{1}{2}\varepsilon_{cd}{}^{ef}\right)\,,
  \,\,\,\,\,\text{with}\,\,\,\,\,
  (\eta_{ab})\equiv \textrm{diag}(-+++)\,,
\end{equation}

\noindent
the spin connection and its curvature 2-form can be rewritten as

\begin{subequations}
\begin{align}
  \omega_{ab}
  &  =
    (\mathbb{N}^{+}_{ab})_{cd}V^{c} v^{d}\,,
  \\
  & \nonumber \\
  R_{ab}
  & = 
    \left\{\tfrac{1}{2} (\mathbb{N}^{+}_{ab})_{ef}V_{c}V^{c}
    +(\mathbb{N}^{+}_{ab})_{eg}(\nabla_{f}V^{g}-V_{f}V^{g})  \right\}
    v^{e}\wedge v^{f}\,,
\end{align}
\end{subequations}

\noindent
with

\begin{equation}
  V_{a} \equiv \partial_{a}\log{H}\,.  
\end{equation}

The matrices $\mathbb{N}^{\pm}_{ab}$ are self- and anti-self-dual in the
indices $ab$, that is,

\begin{equation}
  \mathbb{N}^{\pm}_{ab}
  =
  \pm\tfrac{1}{2}\varepsilon_{abcd}  \mathbb{N}^{\pm}_{cd}\,,
\end{equation}


\noindent
which makes the above spin connection and curvature manifestly self-dual.

Furthermore, the Pontrjagin density can be written as




\begin{equation}
\label{eq:RR}
R^{ab}\wedge R_{ab}  
=
d\star_{(4)}d(\partial\log{H})^{2}\,.
\end{equation}



\noindent
Then, for GH spaces, Eq.~(\ref{eq:forfirstorderterms}) can be brought to the
form

\begin{equation}
  \label{eq:forfirstordertermsGH}
  d\star d \left[\mathcal{H}^{2}\left(C+\phi_{1}^{(-)}\right)
    -\tfrac{1}{8}(\partial\log{H})^{2}  \right]
  =
  0\,.
\end{equation}

\noindent
This equation is then solved by

\begin{equation}
  \phi_{1}^{(-)}
  =
  -C+\frac{1}{\mathcal{H}^{2}}\left(\mathcal{K} +\frac{(\partial\log{H})^{2}}{8}\right)\,,
\end{equation}

\noindent
where $\mathcal{K}$ is a harmonic function in the GH space. At this point,
the following observation becomes relevant: $\eta$-independent harmonic
functions in GH spaces are just harmonic functions in $\mathbb{E}^{3}$.

In summary, we have found that, given a GH space, we can find first-order in
$\alpha'$ solutions of the Euclidean Cano--Ruip\'erez action
Eq.~(\ref{eq:ECRaction3}) which have the same metric (with no $\alpha'$
corrections) and scalars given by

\begin{subequations}
  \label{eq:scalarsinGHspaces}
  \begin{align}
    \phi^{(+)}
    & =
      B +\alpha' C+ \mathcal{O}(\alpha^{\prime\, 2})\,,
    \\
    & \nonumber \\
    \phi^{(-)}
    & =
      -\phi^{(+)} +\frac{1}{\mathcal{H}}
      +\frac{\alpha'}{\mathcal{H}^{2}} \left[\mathcal{K}
      +\frac{(\partial\log{H})^{2}}{8} \right]
      + \mathcal{O}(\alpha^{\prime\, 2})\,,
  \end{align}
\end{subequations}

\noindent
where $B$ and $C$ are arbitrary constants and $\mathcal{H}$ and $\mathcal{K}$
are arbitrary harmonic functions in the GH space, all of them to be suitably
chosen so as to make the scalar fields regular.

If the harmonic function $\mathcal{H}$ has the same general form as $H$ in
Eq.~(\ref{eq:H-GHspaces}) with all constants positive, the singularities
of the scalar fields may only come from the $(\partial\log{H})^{2}$ term,
which is explicitly given by

\begin{equation}
  (\partial\log{H})^{2}
  =
  H^{-3} \sum_{n,m}
    \frac{a_{n}a_{m}(\vec{x}-\vec{x}_{n})\cdot (\vec{x}-\vec{x}_{m})}{|\vec{x}-\vec{x}_{n}|^{3}|\vec{x}-\vec{x}_{m}|^{3}}\,.
\end{equation}

When $\vec{x}\to \vec{x}_{*}$, where $\vec{x}_{*}$ is some particular
$\vec{x}_{n}$, this term diverges as

\begin{equation}
  (\partial\log{H})^{2}
  \sim \frac{1}{a_{*}\varepsilon}\,,
  \hspace{1cm}
  \varepsilon\to 0\,.
\end{equation}

We can cancel these divergences for all $\vec{x}_{n}$, irrespectively of our
choice of $\mathcal{H}$, by choosing $\mathcal{K}$ to be of the
form\footnote{If $\mathcal{H}$ has poles at the same locations as $H$, then
  $(\partial\log{H})^{2}/\mathcal{H}^{2}$ has no divergences to eliminate and
  we should simply be careful not to introduce new divergences with our choice
  of $\mathcal{K}$.}

\begin{equation}
  \label{eq:K-GH}
  \mathcal{K}
  =
  \frac{F}{E^{2}}
  -\sum_{n}\,\frac{1}{8a_{n}|\vec{x}-\vec{x}_{n}|}\,,
\end{equation}

\noindent
for some arbitrary real constant $F$, assuming $\mathcal{H}\sim E$ at
infinity. This choice guarantees that both the dilaton and axion are regular.

\subsection{The Euclidean Taub--NUT space}
\label{sec-ETN}

Perhaps the simplest example of GH space is the so-called Euclidean Taub--NUT
(ETN) space, which corresponds to the choice of harmonic
function\footnote{This name is a bit misleading: the Lorentzian Taub--NUT
  spacetime has two independent parameters: the ADM mass $m$ and the NUT
  charge $n$. The so-called ETN solution corresponds to just to the $m=n$
  Wick-rotated Taub--NUT solution. Another, possibly better, name for the same
  solution is that of Gross-Perry-Sorkin or Kaluza-Klein monopole
  \cite{Gross:1983hb,Sorkin:1983ns}.}

\begin{equation}
H =1+\frac{2n}{|\vec{x}|}\,,  
\end{equation}

\noindent
where $n$ is the NUT charge parameter. In spherical coordinates,
$|\vec{x}|=\rho$ and the 1-form $\chi$ is, locally, given by just
$2n\cos{\theta}d\varphi$. After a change of radial coordinate $\rho=r-n$, the
metric takes the standard form

\begin{equation}\label{eq:ds-TN}
  ds^{2}
  =
  f(r) (d\eta +2n\cos{\theta}d\varphi)^{2}
  +\frac{dr^{2}}{f(r)} +(r^{2}-n^{2})d\Omega^{2}_{(2)}\,,
\end{equation}

\noindent
with

\begin{equation}
  f(r) = \frac{r-n}{r+n}\,,
  \hspace{1cm}
  d\Omega^{2}_{(2)}
  =
  d\theta^{2}+\sin^{2}{\theta}d\varphi^{2}\,,
\end{equation}

\noindent
and it covers the region $r>n$ with $\theta\in [0,\pi]$,
$\varphi \in [0,2\pi)$ and $\eta\in (0,8\pi n]$. The point $r=n$, at which the
metric is completely regular when $\eta\in (0,8\pi n]$ may or not may be
included in the manifold. When it is not, the topology of the manifold is that
of $\mathbb{R}\times S^{3}$ \cite{Eguchi:1977iu}. When it is, the topology is
that of $\mathbb{E}^{4}$ \cite{Hawking:1976jb}.

The main ingredient in the $\alpha'$ correction to the scalar field
$\phi^{(-)}$ is the term

\begin{equation}
  (\partial\log{H})^{2}
  =
  H^{-3}\partial_{\underline{x}}H\partial_{\underline{x}}H
  =
  \frac{4n^{2}}{\rho (\rho+2n)^{3}}
  =
  \frac{4n^{2}}{(r-n)(r+n)^{3}}\,,  
\end{equation}

\noindent
which has a pole at $r=n$ that we can cancel with the specific choice of the
harmonic function $\mathcal{K}$ in Eq.~\eqref{eq:K-GH}, rendering the solution
completely regular at $r=n$. Choosing for simplicity $\mathcal{H}=E^{-1}$, in
the coordinates used in Eq.~\eqref{eq:ds-TN}. $\mathcal{K}$ is given by

\begin{equation}
\mathcal{K}=\frac{F}{E^{2}}-\frac{1}{16n(r-n)}\,,
\end{equation}

\noindent
and we find 

\begin{subequations}
  \label{eq:scalarsinETN}
  \begin{align}
    \phi^{(+)}
    & =
      B +\alpha' C+ \mathcal{O}(\alpha^{\prime\, 2})\,,
    \\
    & \nonumber \\
    \phi^{(-)}
    & =
      -\phi^{(+)} +E+\alpha' \left[F-\frac{E^{2}(r^{2}+4nr+7n^{2})}{16n(r+n)^{3}}\right]
      +\mathcal{O}(\alpha^{\prime\, 2})\,.
  \end{align}
\end{subequations}

In terms of their asymptotic values, $e^{-\phi_{\infty}},\chi_{E\, \infty}$,
the $\alpha^{\prime}$-corrected dilaton and axion fields read

\begin{subequations}
  \label{eq:scalarsinETN2}
  \begin{align}
\chi_{E}
    & =
      \chi_{E\, \infty}
      +\alpha'e^{-2\phi_{\infty}}\frac{(r^{2}+4nr+7n^{2})}{8n(r+n)^{3}}
      +\mathcal{O}(\alpha^{\prime\, 2})\,,
    \\
    & \nonumber \\
     e^{-\phi}
    & =
       e^{-\phi_{\infty}} -\alpha'e^{-2\phi_{\infty}}
       \frac{(r^{2}+4nr+7n^{2})}{8n(r+n)^{3}}
       +\mathcal{O}(\alpha^{\prime\, 2})\,.
  \end{align}
\end{subequations}

\noindent
where

\begin{align}
  e^{-\phi_{\infty}} = \tfrac{1}{2}(E+\alpha'F)\,,
  \qquad \text{and} \qquad
  \chi_{E\, \infty} =   B+\alpha'C -e^{-\phi_{\infty}}\,.
\end{align}

\noindent
However, other no-trivial choices of the harmonic function $\mathcal{H}$ are also possible.

\subsection{The Eguchi--Hanson instanton}
\label{sec-EguchiHanson}

The Eguchi--Hanson instanton~\cite{Eguchi:1978xp,Eguchi:1978gw} is the GH space corresponding
to the two-instanton solution. It can be obtained by choosing of harmonic function as~\cite{Prasad:1979kg}

\begin{equation}
  H
  =
  \frac{1}{|\vec{x}_{+}|}
  +\frac{1}{|\vec{x}_{-}|}\,,
  \hspace{1cm}
  \vec{x}_{\pm} \equiv (x^{1},x^{2},x^{3}_{\pm})\,,
  \hspace{1cm}
  x^{3}_{\pm} \equiv x^{3}\pm x^{3}_{0}\,,
\end{equation}

\noindent
but it is usually presented in the form

\begin{equation}
  ds^{2}
  =
  \frac{r^{2}}{4}f(r) [d\psi+\cos{\theta}d\varphi]^{2}
  +\frac{dr^{2}}{f(r)} 
  +\frac{r^{2}}{4} [d\theta^{2}+\sin^{2}{\theta}d\varphi^{2}]\,,
  \hspace{1cm}
      f(r) = 1 - \left(\frac{r_{0}}{r}\right)^{4}\,.
\end{equation}

The change of coordinates that brings the metric to this form, found by Prasad
in Ref.~\cite{Prasad:1979kg}, is most conveniently made in two steps: first,
one transforms the coordinates $x^{x}$ into elliptical coordinates
$\alpha,\beta,\gamma$

\begin{equation}
  \begin{aligned}
    x^{1}
    & = x^{3}_{0}\sinh{\alpha}\sin{\beta}\cos{\gamma}\,,
      \\
    x^{2}
    & = x^{3}_{0}\sinh{\alpha}\sin{\beta}\sin{\gamma}\,,
      \\
    x^{3}
    & = x^{3}_{0}\cosh{\alpha}\cos{\beta}\,,
  \end{aligned}
\end{equation}

\noindent
and, then, one identifies

\begin{align}
    \beta
    & =
      \theta\,,
    &
    \eta
    & =
      2\varphi\,,
    &
    \gamma
    & =
      \psi\,,
      &
    8x^{3}_{0}
    & =
      r_{0}^{2}\,,
    &
    8x^{3}_{0}\cosh{\alpha}
    & =
      r^{2}\,.
  \end{align}

The geometrical and topological properties of this manifold have been studied
thoroughly in Refs.~\cite{Eguchi:1978xp,Eguchi:1978gw}. Let us mention some of
them. First, notice that the potential conical singularity at $r=r_{0}$ is
removed by a suitable periodicity of the Euler angle $\psi$, namely
$\psi\sim\psi+2\pi$. Then, the Eguchi--Hanson metric is regular for
$r_{0}<r<\infty$, $0\leq\vartheta\leq\pi$, $0\leq\varphi<2\pi$, and
$0\leq\psi\leq 2\pi$. This, in its turn, implies that the topology near
$r=r_{0}$ is that of $\mathbb{R}^{2}\times S^{2}$. Additionally, it is
asymptotically locally, but not globally, Euclidean as its boundary is the
lens space $S^{3}/\mathbb{Z}_{2} = \mathbb{RP}^{3}$. Its Euler characteristic
is $\chi=2$, and its Hirzebruch signature is $\tau=-1$. The index of the Dirac
operator is $I_{1/2}=0$, but that of the Rarita-Schwinger operator is
$I_{3/2}=-2$. Generalizations in the presence of the cosmological constant,
Maxwell fields, and higher-curvature corrections have been studied in
Refs.~\cite{Eguchi:1978gw,Pedersen:1985,Corral:2022udb,Corral:2025yvr}
and it has been used as a seed metric for constructing solitons in five
dimensions in
Refs.~\cite{Clarkson:2005qx,Clarkson:2006zk,Chng:2006gh,Wong:2011aa,Durgut:2022xzw}.

The $\alpha^{\prime}$ corrections to the dilaton and axion over the
Eguchi--Hanson instanton can be easily found using the original GH
coordinates. Taking into account the choice of the harmonic functions
$\mathcal{H}=E^{-1}$ and $\mathcal{K}$ in Eq.~\eqref{eq:K-GH} that renders the
solution regular at $r=r_{0}$, we find

\begin{align}
  \chi_{E}
  & =
    \chi_{E\,\infty}
    +\alpha^{\prime} e^{-2\phi_{\infty}}\,
    \frac{2\left(3r^{4} + r_{0}^{4} \right)}{r^{6}} + \mathcal{O}(\alpha^{\prime\, 2})\,, \\
  &\nonumber \\
  e^{-\phi}
  & =
    e^{-\phi_{\infty}}
    -\alpha^{\prime}e^{-2\phi_\infty}
    \frac{2\left(3r^{4} + r_{0}^{4} \right)}{r^{6}}+ \mathcal{O}(\alpha^{\prime\, 2})\,,
\end{align}

\noindent
where $\phi_{\infty}$ is the asymptotic value of the dilaton. The nontrivial
nature of this configuration relies critically on the stringy corrections
since the dilaton and axion become constant when $\alpha'\to 0$.

\section{The Euclidean action}
\label{sec-euclideanaction}

As we mentioned in Section~\ref{sec-thetheory}, the Euclidean
Cano--Ruip\'erez action Eq.~(\ref{eq:ECRaction3}) has to be
supplemented by a boundary term such that

\begin{enumerate}
\item Its variation cancels the boundary term in the variation of the
  volume term assuming Dirichlet-type boundary conditions for the
  variations of the fields (leaving the derivatives of the variations
  unconstrained).
\item The total action vanishes for a chosen vacuum solution.
\end{enumerate}

It is convenient to start by reviewing the well-known Gibbons--Hawking--York
boundary term of pure Einstein gravity~\cite{York:1972sj,Gibbons:1976ue}. This
will help us to establish the notation we are going to use and the logic
behind the construction. Then, we will consider boundary terms for the
Gauss--Bonnet and Pontrjagin densities as intermediate steps before
considering the boundary term of the whole action. We will ignore the overall
normalization coefficient of the action throughout, and we will only consider
the Euclidean signature case.

\subsection{The Gibbons--Hawking--York boundary term}
\label{app-boundarytermGHY}

The variation of the Einstein-Hilbert term in the action

\begin{equation}
\int_{M}\varepsilon_{abcd}e^{a}\wedge e^{b}\wedge R^{cd}\,,
\end{equation}

\noindent
gives, after use of the Stokes therm, a boundary term

\begin{equation}
\int_{\partial M}\varepsilon_{abcd}e^{a}\wedge e^{b}\wedge \delta \omega^{cd}\,,  
\end{equation}

\noindent
that contains variations of the derivatives of the Vierbein. Naively, we could
cancel this term by adding to the Einstein-Hilbert action a boundary
term\footnote{The terms proportional to the variations of the Vierbein vanish
  with the standard Dirichlet-type boundary conditions
  $\left. \delta e^{a}\right|_{\partial M}=0$.}

\begin{equation}
  \label{eq:naiveGHYterm}
-\int_{\partial M}\varepsilon_{abcd}e^{a}\wedge e^{b}\wedge \omega^{cd}\,,  
\end{equation}

\noindent
but this term would break local Lorentz invariance because
$\omega^{cd}$ is a connection. Since the difference between two connections transform as a tensor, we can construct a well-defined
boundary term by replacing $\omega^{ab}$ by the difference
$\omega^{ab}-\omega_{0}{}^{ab}$, where $\omega_{0}{}^{ab}$ is a fixed
connection whose variation vanishes, that is

\begin{equation}
  \label{eq:naiveGHYterm2}
  -\int_{\partial M}\varepsilon_{abcd}e^{a}\wedge e^{b}\wedge
  \left(\omega^{cd}-\omega_{0}{}^{cd}\right)\,.  
\end{equation}

The standard prescription for $\omega_{0}^{ab}$ in the computation of
boundary terms of topological invariants is to use the Levi--Civita
connection of a product metric $e_{0}^{a}$ which coincides with the
metric under consideration $e^{a}$ on the boundary. This prescription
may require the use of different metrics $e_{0}^{a}$ in different,
disconnected pieces of the boundary. The difference

\begin{equation}
  \omega^{ab}-\omega_{0}{}^{ab}
  \equiv
  \vartheta^{ab}\,,
\end{equation}

\noindent
is the \textit{second fundamental form} of the boundary and only has
non-vanishing components in the components that are tangent to the
boundary.

Indeed, if we choose a Vierbein basis in which
$e^{\flat}=\mathbf{n}=n_{\mu}dx^{\mu}$, the 1-form dual to the unit
vector $n=n^{\mu}\partial_{\mu}$ normal to the boundary and we call
$e^{i}$, $i=1,2,3$, the other three 1-forms of the basis, a
straightforward calculation using Eq.~(\ref{eq:KLne}) leads to

\begin{equation}
  \label{eq:Kversusvartheta}
  \vartheta^{i}{}_{\flat}
  =
 \mathcal{K}^{i}{}_{j}e^{j}\,,
  \hspace{1cm}
  \vartheta^{i}{}_{j} =0\,,
\end{equation}

\noindent
where $\mathcal{K}^{i}{}_{j}$ are the tangent components of the
\textit{extrinsic curvature} tensor (or second fundamental form).

For the sake of completeness, we remind the reader that, if 

\begin{equation}
  h_{\mu\nu}
  \equiv
  g_{\mu\nu}-n_{\mu}n_{\nu}\,,
\end{equation}

\noindent
is the induced metric on a surface with unit normal vector $n$, the extrinsic
curvature tensor is defined as

\begin{equation}
  \label{eq:Kmunudef}
  \begin{aligned}
    \mathcal{K}_{\mu\nu}
    & \equiv
    \tfrac{1}{2}
      \mathcal{L}_{n}h_{\mu\nu}
    \\
    & \\
    & =
      \nabla_{(\mu}n_{\nu)} -n_{(\mu|}n^{\alpha}\nabla_{\alpha}n_{|\nu)}
    \\
    & \\
    & =
      \nabla_{\mu}n_{\nu} -n_{\mu}n^{\alpha}\nabla_{\alpha}n_{\nu}\,,
  \end{aligned}
\end{equation}

\noindent
where the last equality follows from the condition of hypersurface
orthogonality

\begin{equation}
  n_{[\mu}\nabla_{\nu}n_{\rho]}=0\,.
\end{equation}

It can be seen that

\begin{equation}
  \mathcal{K}_{\mu\nu}
  =
  h_{\mu}{}^{\alpha}h_{\nu}{}^{\beta} \nabla_{\alpha}n_{\beta}\,,
  \hspace{1cm}
n^{\mu}  \mathcal{K}_{\mu\nu}
=
0\,,
\end{equation}

\noindent
which shows that it only has tangent components.

In the above Vierbein basis

\begin{equation}
  h_{\mu\nu}dx^{\mu}dx^{\nu}
  =
  e^{i} e^{i}\,,
\end{equation}

\noindent
and it is not difficult to show that the definition of
$\mathcal{K}_{\mu\nu}$ Eq.~\eqref{eq:Kmunudef} leads to

\begin{equation}
  \label{eq:KLne}
\mathcal{L}_{n}e^{i} = \mathcal{K}^{i}{}_{j}e^{j}\,,  
\end{equation}

\noindent
from which Eqs.~\eqref{eq:Kversusvartheta} follow immediately.

Replacing the connection by the second fundamental form in the
boundary term Eq.~\eqref{eq:naiveGHYterm} we obtain

\begin{equation}
  \label{eq:GHYdiffform1}
  -\int_{\partial M}\varepsilon_{abcd}e^{a}\wedge e^{b}\wedge \vartheta^{cd}\,,  
\end{equation}

\noindent
which is exactly Lorentz invariant and whose variation cancels the
total derivative term of the Einstein--Hilbert term with the standard
boundary conditions. Using Eqs.~\eqref{eq:Kversusvartheta} and
assuming the orientation $\varepsilon^{123\flat}=+1$ we can rewrite
this term in the more standard form

\begin{equation}
  \label{eq:GHYterm}
    -2\int_{\partial M}\varepsilon_{ijk}e^{i}\wedge e^{j}\wedge e^{l}
    \mathcal{K}^{k}{}_{l}
    =
    -2\int_{\partial M}d^{3}x\,\sqrt{|h|}\varepsilon_{ijk}\varepsilon^{ijl}
    \mathcal{K}^{k}{}_{l}
    =
    -4\int_{\partial M}d^{3}x\,\sqrt{|h|}\,\mathcal{K} \,,  
\end{equation}

\noindent
where $|h|$ is the determinant of the metric induced on the boundary and

\begin{equation}
  \mathcal{K}
  \equiv
  \mathcal{K}^{\mu}{}_{\mu}\,,
\end{equation}

\noindent
is the trace of the extrinsic curvature.

This is the boundary term first proposed by York in
Ref.~\cite{York:1972sj}.  In Ref.~\cite{Gibbons:1976ue}, Gibbons and
Hawking proposed to replace $\mathcal{K}$ in the above form of the
boundary term by $\mathcal{K}-\mathcal{K}_{B}$, \textit{i.e.}

\begin{equation}
  \label{eq:GHYtensorform} 
-4\int_{\partial M}d^{3}x\,\sqrt{|h|}\,\left[\mathcal{K}-\mathcal{K}_{B}\right]\,,  
\end{equation}

\noindent
where $\mathcal{K}_{B}$ is obtained from $\mathcal{K}$ by using in it a
reference, background metric, which is treated as the ground state. The above
term (the GHY boundary term) vanishes identically for the background
metric. This was, actually, the main motivation for the subtraction of
$\mathcal{K}_{B}$. Most of the cases considered in Ref.~\cite{Gibbons:1976ue}
are asymptotically-flat spaces for which the most appropriate choice of ground
state is flat, Euclidean space, for which the volume term in the action also
vanishes.

The analog of Eq.~\eqref{eq:GHYtensorform} in the differential-form
language used in Eq.~\eqref{eq:GHYdiffform} is

\begin{equation}
  \label{eq:GHYdiffform}
  -\int_{\partial M}\varepsilon_{abcd}e^{a}\wedge e^{b}\wedge
  \left[\vartheta^{cd}-\vartheta_{B}^{cd}\right]\,.
\end{equation}

\noindent
Notice, though, that this term is not fully identical to
Eq.~\eqref{eq:GHYtensorform} when the metric on the boundary does not coincide
with the background metric, but it has the desired property of vanishing for
the background metric.

The above boundary term reproduces as closely as possible the GHY
boundary term. However, in differential-form language this is not the
only possible boundary term with the desired properties: as we have
discussed at the beginning of this section, the naive boundary term
Eq.~\eqref{eq:naiveGHYterm2} is a Lorentz scalar whose variation
cancels the boundary term in the variation of the Einstein--Hilbert
action for any fixed connection $\omega_{0}{}^{cd}$ and we can make it
vanish for the chosen background by choosing
$\omega_{0}{}^{cd}=\omega_{B}{}^{cd}$.

It is not difficult to see that

\begin{equation}
  \label{eq:GHYdiffformfinal}
  -\int_{\partial M}\varepsilon_{abcd}e^{a}\wedge e^{b}\wedge
  \vartheta_{*}{}^{cd}\,,
  \hspace{1cm}
  \vartheta_{*}{}^{cd}
  \equiv
  \omega^{cd}-\omega_{B}{}^{cd}\,,  
\end{equation}

\noindent
gives the same result for the action of the Euclidean Schwarzschild
solution, for instance. Thus, for the sake of simplicity, in what
follows we will use the 1-form $\vartheta_{*}{}^{ab}$ instead of the
difference between the second fundamental form and the second
fundamental form evaluated over the background.

\subsection{The boundary term of the Gauss--Bonnet density}
\label{app-boundarytermGB}

Let us define the 4-dimensional Gauss--Bonnet density as

\begin{equation}
  \mathbf{GB}
  \equiv
  2\tilde{R}^{ab}\wedge R_{ab}
  =
  \varepsilon_{abcd} R^{ab}\wedge R^{cd}\,,
\end{equation}

\noindent
and consider an action whose volume term is the integral if this density. This
density is a total derivative that we can write in the form\footnote{The proof
  of this fact makes use of the Schoutens-type identities
  \begin{equation}
    0
    =
    5\varepsilon_{[abcd}g_{e]f}X^{ab}\wedge \omega^{cd}\wedge \omega^{ef}
    =
    2\varepsilon_{abcd}\left[
      X^{ab}\wedge \omega^{c}{}_{e}\wedge \omega^{ed}
    +\omega^{ab}\wedge X^{c}{}_{e}\wedge \omega^{ed}\right]\,,
\end{equation}
for any 2-form $X^{ab}=X^{[ab]}$, setting $X^{ab}=d\omega^{ab}$ and also
$X^{ab}=\omega^{a}{}_{c}\wedge \omega^{cb}$.
}

\begin{subequations}
  \begin{align}
  \mathbf{GB}
  & =
  d\left\{\varepsilon_{abcd}\left[d\omega^{ab}\wedge \omega^{cd}
    -\tfrac{2}{3}\omega^{ab}\wedge \omega^{c}{}_{e}\wedge \omega^{ed} \right]
    \right\}
    \\
    & \nonumber \\
    & =
        d\left\{\varepsilon_{abcd}\left[R^{ab}\wedge \omega^{cd}
    +\tfrac{1}{3}\omega^{ab}\wedge \omega^{c}{}_{e}\wedge \omega^{ed} \right]
    \right\}\,,
  \end{align}
\end{subequations}

\noindent
and, under arbitrary variations of the connection, it transforms as

\begin{equation}
  \delta   \mathbf{GB}
  =
  d\left[4 \tilde{R}_{ab}\wedge \delta \omega^{ab} \right]\,.
\end{equation}

\noindent
Although this action does not lead to non-trivial equations of motion, its
variation leads to a boundary term that contains derivatives of the variations
of the Vierbein, which should be canceled by an appropriate boundary term. It
is worth mentioning that, in Einstein-AdS gravity, the Gauss--Bonnet term can
be used as a counterterm that defines a well-posed variational principle for
the boundary metric with Dirichlet boundary
conditions~\cite{Aros:1999id,Olea:2005gb,Anastasiou:2020zwc}, while the
Pontrjagin term allows one to set all self-dual configurations as the ground
state of the theory~\cite{Miskovic:2009bm,Araneda:2016iiy,Corral:2024lva}.

A 3-form whose variation gives exactly the same boundary term when evaluated
over the boundary, and which vanishes identically when evaluated over the
chosen background, is

\begin{equation}\label{GBb}
  \begin{aligned}
  \mathbf{GB}_{b}
  & \equiv
  \varepsilon_{abcd}
  \left\{\left(R^{ab}+R_{B}{}^{ab}\right)\wedge \vartheta_{*}{}^{cd}
    +\tfrac{1}{3}\vartheta_{*}{}^{ab}\wedge
    \vartheta_{*}{}^{c}{}_{e}\wedge \vartheta_{*}{}^{ed} \right\}\,,
  \end{aligned}
\end{equation}

\noindent
where $R_{B}{}^{ab}$ is the curvature of $\omega_{B}{}^{ab}$; the spin
connection of the fixed background metric, satisfying $\delta \omega_{B}{}^{ab}=0$. Additionally, $\vartheta_{*}{}^{ab}$, defined in Eq.~\eqref{eq:GHYdiffformfinal}, vanishes for the background
metric. Using the identity

\begin{equation}
  \label{eq:identity}
  \mathcal{D}\vartheta_{*}{}^{a}{}_{b}
  =
  R^{a}{}_{b}-R_{B}{}^{a}{}_{b}
  -\vartheta_{*}{}^{a}{}_{c}\wedge \vartheta_{*}{}^{c}{}_{b}\,,
\end{equation}

\noindent
it is not difficult to see that, over the boundary, the variation of Eq.~\eqref{GBb} yields

\begin{equation}
    \label{eq:deltaGBb}
  \delta \mathbf{GB}_{b}
  =
  4 \tilde{R}_{ab}\wedge \delta \omega^{ab}
  +d\left(-\varepsilon_{abcd}\delta\omega^{ab}\wedge\omega_{B}{}^{cd}\right)\,,
\end{equation}

\noindent
so that

\begin{equation}
  \delta \left\{\int_{M}\mathbf{GB} -\int_{\partial M}\mathbf{GB}_{b} \right\}
  =
  0\,.
\end{equation}



If our background metric is not flat, the volume term may not be zero
for it. We may have to subtract from the volume term the
Gauss--Bonnet density of the background metric, if we want to normalize
the complete action to zero for it.

\subsection{The boundary term of the Pontrjagin density}
\label{app-boundarytermP}

The 4-dimensional Pontrjagin density can be defined as

\begin{equation}
  \mathbf{P}
  \equiv
  R^{a}{}_{b}\wedge R^{b}{}_{a}\,.
\end{equation}

\noindent
It is the total derivative of the Chern-Simons 3-form, that is,

\begin{subequations}
  \begin{align}
    \mathbf{P}
    & =
      d\left\{d\omega^{a}{}_{b}\wedge \omega^{b}{}_{a}
      -\tfrac{2}{3}\omega^{a}{}_{b}\wedge \omega^{b}{}_{c}\wedge
      \omega^{c}{}_{a}\right\}
    \\
    & \nonumber \\
    & =
      d\left\{R^{a}{}_{b}\wedge \omega^{b}{}_{a}
      +\tfrac{1}{3}\omega^{a}{}_{b}\wedge \omega^{b}{}_{c}\wedge
      \omega^{c}{}_{a}\right\}\,,    
  \end{align}
\end{subequations}

\noindent
and, under arbitrary variations of the connection, it gives

\begin{equation}
\delta    \mathbf{P}
=
d\left[2 R^{a}{}_{b}\wedge \delta\omega^{b}{}_{a} \right]\,.
\end{equation}

Again, an action given by the integral of this density needs a
boundary term whose variation cancels the above term.

A 3-form whose variation gives the same boundary term (up to a total
derivative) when evaluated over the boundary and which also vanishes when
$\omega^{ab}=\omega_{B}{}^{ab}$ is

\begin{equation}
  \begin{aligned}
  \mathbf{P}_{b}
  & \equiv
  \left(R^{a}{}_{b}+R_{B}{}^{a}{}_{b} \right)\wedge \vartheta_{*}{}^{b}{}_{a}
  +\tfrac{1}{3}\vartheta_{*}{}^{a}{}_{b}\wedge \vartheta_{*}{}^{b}{}_{c}\wedge
    \vartheta_{*}{}^{c}{}_{a}\,.
  \end{aligned}
\end{equation}

\noindent
We have, in fact,

\begin{equation}
  \delta \mathbf{P}_{b}
  =
  2R^{a}{}_{b}\wedge \delta\omega^{b}{}_{a}
   +d\left(-\delta\omega^{a}{}_{b}\wedge\omega_{B}{}^{b}{}_{a}\right)\,,
\end{equation}

\noindent
and, therefore,

\begin{equation}
  \label{eq:deltaPb}
  \delta \left\{\int_{M}\mathbf{P} -\int_{\partial M}\mathbf{P}_{b} \right\}
  =
  0\,.
\end{equation}

Again, using the identity Eq.~(\ref{eq:identity}), we can rewrite the boundary
term in the form 

\begin{equation}
  \mathbf{P}_{b}
  =
  2R^{a}{}_{b}\wedge \vartheta_{*}{}^{b}{}_{a}
  -\tfrac{2}{3}\vartheta_{*}{}^{a}{}_{b}\wedge \vartheta_{*}{}^{b}{}_{c}\wedge
  \vartheta_{*}{}^{c}{}_{a}
  +2\omega^{a}{}_{c}\wedge \vartheta_{*}{}^{c}{}_{b}\wedge \vartheta_{*}{}^{b}{}_{a}
    -d\vartheta_{*}{}^{a}{}_{b}\wedge \vartheta_{*}{}^{b}{}_{a}\,.
\end{equation}

Similar to the Gauss--Bonnet case, the use of a non-flat metric as background
may require the subtraction of the Pontrjagin density of the background from
the volume term for the complete action to be properly normalized.

\subsection{The complete boundary term of the Euclidean Cano--Ruip\'erez action}
\label{app-totalboundarytermECR}

Equipped with this knowledge, we can now propose a boundary term for the
Euclidean Cano--Ruip\'erez action for Dirichlet-type boundary conditions,
\textit{i.e.}~$\delta e^{a} = \delta \phi = \delta \chi_{E} = 0$ on
$\partial M$. First, notice that the boundary term must cancel the
presymplectic potential $\mathbf{\Theta}(\varphi,\delta\varphi)$ in
Eq.~(\ref{eq:presymplecticpotential}) with those boundary conditions, namely

\begin{equation}
  \begin{aligned}
    \left.\mathbf{\Theta}(\varphi,  \delta \varphi)\right|_{\partial M}
     & =
     \left[\star (e^{a}\wedge e^{b}) +\mathbf{H}_{ab}\right]
       \wedge \delta \omega_{ab}
    \\
     & \\
    & =
      \star (e^{a}\wedge e^{b})\wedge \delta \omega_{ab}
      -\frac{\alpha'}{2}e^{-\phi}\tilde{R}^{ab}\wedge \delta \omega_{ab}
    -\frac{\alpha'}{2}\chi_{E}R^{ab} \wedge \delta \omega_{ab}\,.
  \end{aligned}
\end{equation}

Assuming Dirichlet-type boundary conditions,
$\delta e^{a} = \delta \phi = \delta \chi_{E} = 0$ on
$\partial M$, the variations of the boundary term

\begin{equation}
-\mathbf{B}
  \equiv
  \star (e^{a}\wedge e^{b})\wedge \vartheta_{*}{}_{ab}
      -\frac{\alpha'}{8}e^{-\phi}\mathbf{GB}_{b}
    +\frac{\alpha'}{4}\chi_{E}\mathbf{P}_{b}\,,
\end{equation}

\noindent
which vanishes identically for $\omega^{ab}=\omega_{B}{}^{ab}$ only
depend on $\delta \omega^{ab}$ and are given by

\begin{equation}
  \begin{aligned}
-\left.\delta  \mathbf{B} \right|_{\partial M}
    & =
  \star (e^{a}\wedge e^{b})\wedge \delta \omega_{ab}
      -\frac{\alpha'}{8}e^{-\phi}\delta\mathbf{GB}_{b}
    +\frac{\alpha'}{4}\chi_{E}\delta\mathbf{P}_{b}
    \\
    & \\
    & =
    \left.\mathbf{\Theta}(\varphi,  \delta \varphi)\right|_{\partial M}
      +\frac{\alpha'}{8}e^{-\phi}
      d\left(\varepsilon_{abcd}\delta\omega^{ab}\wedge\omega_{B}{}^{cd}\right)
      +\frac{\alpha'}{4}\chi_{E}
      d\left(\delta\omega^{a}{}_{b}\wedge\omega_{B}{}^{b}{}_{a}\right)
    \\
    & \\
    & =
    \left.\mathbf{\Theta}(\varphi,  \delta \varphi)\right|_{\partial M}
      -\frac{\alpha'}{8}\varepsilon_{abcd} de^{-\phi}\wedge 
      \delta\omega^{ab}\wedge\omega_{B}{}^{cd}
      -\frac{\alpha'}{4}d\chi_{E}\wedge 
      \delta\omega^{a}{}_{b}\wedge\omega_{B}{}^{b}{}_{a}
    \\
    & \\
    & \hspace{.5cm}
      +d(\cdots)\,,
  \end{aligned}
\end{equation}

\noindent
where we have used Eqs.~(\ref{eq:deltaGBb}) and (\ref{eq:deltaPb}) and we have
integrated by parts. The second and third term vanish over the boundary if we
assume that the scalars approach a constant value over the boundary so that
$d\phi=d\chi_{E}=0$ there.

Thus, under these assumptions,

\begin{equation}
  \delta \left\{S_{ECR}+\int_{\partial M}\mathbf{B}\right\}
    =
          \int_{M}
      \left\{
      \mathbf{E}_{a} \wedge \delta e^{a}
      +\mathbf{E}_{(+)} \, \delta \phi^{(+)}
      +\mathbf{E}_{(-)} \, \delta \phi^{(-)}
      \right\}\,.
\end{equation}

The action will not vanish for the background solution if that
solution is not flat spacetime with constant scalars. In that case we
may have to subtract some of the volume terms evaluated over the
background solution. We will avoid this complication in what follows
restricting ourselves to asymptotically-flat solutions for which the
background metric can safely be taken as flat space $R_{B}^{ab}=0$.

The boundary term can be rewritten in the form

\begin{equation}
  \label{eq:boundarytermplusminus}
  \begin{aligned}
\mathbf{B}
  & =
  -\star (e^{a}\wedge e^{b})\wedge \vartheta_{*}{}_{ab}
      +\frac{\alpha'}{4}\phi^{(+)}\left\{R^{(+)\, ab}\wedge \vartheta_{*}{}_{ab}
    +\tfrac{1}{3}\vartheta_{*}{}^{(+)\,  ab}\wedge
    \vartheta_{*}{}_{ae}\wedge \vartheta_{*}{}^{e}{}_{b}\right\}
    \\
    & \\
    & \hspace{.5cm}
    -\frac{\alpha'}{4}\phi^{(-)}\left\{R^{(-)\, ab}\wedge \vartheta_{*}{}_{ab}
    +\tfrac{1}{3}\vartheta_{*}{}^{(-)\,  ab}\wedge
      \vartheta_{*}{}_{ae}\wedge \vartheta_{*}{}^{e}{}_{b}\right\}\,,
  \end{aligned}
\end{equation}

\noindent
which is more appropriate to deal with field configurations with self- or
anti-self-dual curvatures.



Finally, the complete Euclidean Cano--Ruip\'erez action, including the
boundary term Eq.~(\ref{eq:boundarytermplusminus}) and the
normalization factors is

\begin{equation}
  \begin{aligned}
  S_{\rm ECR}[e^{a},\phi^{(+)},\phi^{(-)}]
  & =
    \frac{1}{16\pi G_{N}} \int_{M} \left\{ \star(e^{a}\wedge e^{b}) \wedge R_{ab}
    -2 \frac{d\phi^{(+)} \wedge \star d\phi^{(-)}}{\left[\phi^{(+)}+\phi^{(-)}\right]^{2}}
    \right.
  \\
    & \\
  & \hspace{.5cm}
    \left.
    -\frac{\alpha'}{4}\left[\phi^{(+)}R^{(+)\, ab}\wedge R^{(+)}{}_{ab}
    -\phi^{(-)}R^{(-)\, ab}\wedge R^{(-)}{}_{ab}\right] \right\}
    \\
    & \\
    &  \hspace{.5cm}
      +\frac{1}{16\pi G_{N}}
      \int_{\partial M}\left\{  -\star (e^{a}\wedge e^{b})\wedge
      \vartheta_{*\, ab}
      \right.
    \\
    & \\
    & \hspace{.5cm}
      +\frac{\alpha'}{4}\phi^{(+)}\left[
R^{(+)\, ab}\wedge \vartheta_{*\, ab}
    +\tfrac{1}{3}\vartheta_{*}{}^{(+)\,  ab}\wedge
    \vartheta_{*\, ae}\wedge \vartheta_{*}{}^{e}{}_{b}\right]
    \\
    & \\
    & \hspace{.5cm}
      \left.
      -\frac{\alpha'}{4}\phi^{(-)}\left[
      R^{(-)\, ab}\wedge \vartheta_{*\, ab}
    +\tfrac{1}{3}\vartheta_{*}{}^{(-)\,  ab}\wedge
      \vartheta_{*\, ae}\wedge \vartheta_{*}{}^{e}{}_{b}
      \right]
      \right\}\,.
  \end{aligned}
\end{equation}

Evaluated on field configurations with self-dual curvature

\begin{equation}
  R^{(-)\, ab}
  =
  0\,,
  \hspace{1cm}
    R^{(+)\, ab}
  =
  R^{ab}\,,
    \hspace{1cm}
    R_{\rm ic}
    =
    0\,,
\end{equation}

\noindent
and with a constant $\phi^{(+)}$, it takes a much simpler
form:

\begin{equation}
  \label{eq:onshellECRactionwithboundaryterms}
  \begin{aligned}
16\pi G_{N}  S_{\rm ECR}^{(+)}
    & =
    -\frac{\alpha' \phi^{(+)}}{4} \int_{M} R^{ab}\wedge R_{ab}
    \\
    & \\
    &  \hspace{.5cm}
      +\int_{\partial M}\left\{  -\star (e^{a}\wedge e^{b})\wedge
      \vartheta_{*\, ab}
      \right.
    \\
    & \\
    & \hspace{.5cm}
      +\frac{\alpha'}{4}\phi^{(+)}\left[
R^{ab}\wedge \vartheta_{*\, ab}
    +\tfrac{1}{3}\vartheta_{*}{}^{(+)\,  ab}\wedge
    \vartheta_{*\, ae}\wedge \vartheta_{*}{}^{e}{}_{b}\right]
    \\
    & \\
    & \hspace{.5cm}
      \left.
      -\frac{\alpha'}{4}\phi^{(-)}\left[
    \tfrac{1}{3}\vartheta_{*}{}^{(-)\,  ab}\wedge
      \vartheta_{*\, ae}\wedge \vartheta_{*}{}^{e}{}_{b}
      \right]
      \right\}\,.
  \end{aligned}
\end{equation}

We now proceed to a detailed evaluation of the action for the only one of the
two particular examples considered before that is asymptotically flat.

\subsection{The Euclidean action of the  $\alpha'$-corrected Eguchi--Hanson metric}
\label{sec-EHaction}

It is convenient to use the Vielbein basis 

\begin{equation}
  \label{eq:VierbeinbasisEH}
  \begin{array}{rclrcl}
e^{1}& = & f^{-1/2}dr\,,\hspace{3cm} &  e^{2} & = & \tfrac{1}{2}rv_{L}^{1}\,, \\
    & & & & & \\
e^{3} & = & \tfrac{1}{2}rv_{L}^{2}\,, &  e^{\sharp} & = & \tfrac{1}{2}rf^{1/2}v_{L}^{3}\,, \\
  \end{array}
\end{equation}

\noindent
where the $v_{L}^{i}$ are the left-invariant Maurer--Cartan 1-forms of
the $SU(2)$ group. They are given by

\begin{equation}
  \label{eq:vLbasis}
  \begin{aligned}
    v_{L}^{1}
    & =
      -\sin{\psi}d\theta +\sin{\theta}\cos{\psi}d\varphi\,,
    \\
    & \\
    v_{L}^{2}
    & =
\cos{\psi}d\theta +\sin{\theta}\sin{\psi}d\varphi\,,
    \\
    & \\
    v_{L}^{3}
    & =
d\psi +\cos{\theta}d\varphi\,,
  \end{aligned}
\end{equation}

\noindent
and satisfy the Maurer-Cartan equations

\begin{equation}
dv_{L}^{i} = \tfrac{1}{2}\varepsilon^{ijk}v_{L}^{j}\wedge v_{L}^{k}\,.
\end{equation}

\noindent
Notice that the volume form of the 3-sphere is given by
$-\tfrac{1}{8}v_{L}^{1}\wedge v_{L}^{2} \wedge v_{L}^{3}$ with the
orientation $\varepsilon^{\theta\varphi\psi}=+1$, so the volume
integration measure on the 3-sphere, $\omega_{(3)}$ is

\begin{equation}
  \tfrac{1}{8}v_{L}^{1}\wedge v_{L}^{2} \wedge v_{L}^{3}
  =
  -\tfrac{1}{8}\sin{\theta}d\theta \wedge d\varphi \wedge d\psi
  \equiv
  -\omega_{(3)}\,,
  \hspace{1cm}
  \int_{S^{3}}\omega_{(3)}
  =
  2\pi^{2}\,.
\end{equation}

In this basis, the components of the spin connection are self-dual with the
orientation $\varepsilon^{\sharp 123}=+1$ in the Vielbein basis and
$\varepsilon^{r\theta\varphi\psi}=+1$ in the coordinate basis

\begin{equation}
  \begin{aligned}
    \omega^{\sharp 1}
    & =
      \omega^{23}
      =
      \tfrac{1}{2}(f-2)v_{L}^{3}\,,
    \\
    & \\
    \omega^{\sharp 2}
    & =
      -\omega^{13}
      =
      -\tfrac{1}{2}f^{1/2}v_{L}^{2}\,,
    \\
    & \\
    \omega^{\sharp 3}
    & =
      \omega^{12}
      =
      \tfrac{1}{2}f^{1/2}v_{L}^{1}\,,    
  \end{aligned}
\end{equation}

\noindent
and so is the curvature

\begin{equation}
  \begin{aligned}
    R^{\sharp 1}
    & =
      R^{23}
      =
      -4\frac{r_{0}^{4}}{r^{6}}(e^{\sharp}\wedge e^{1}+e^{2}\wedge e^{3})\,,
    \\
    & \\
    R^{\sharp 2}
    & =
      -R^{13}
      =
      2\frac{r_{0}^{4}}{r^{6}}(e^{\sharp}\wedge e^{2}-e^{1}\wedge e^{3})\,,
    \\
    & \\
    R^{\sharp 3}
    & =
      R^{12}
      =
      2\frac{r_{0}^{4}}{r^{6}}(e^{\sharp}\wedge e^{3}+e^{1}\wedge e^{2})\,.
  \end{aligned}
\end{equation}

The background solution in this case is just given by the leading
terms in the $r\to\infty$ limit expansion of the $\alpha'$-corrected
solution, whose metric is locally that of flat space (with
$\psi\in[0,2\pi)$ instead of $\psi\in[0,4\pi)$) and constant scalars
$\phi^{(\pm)}=\phi^{(\pm)}_{\infty}$.  Its spin connection, also
self-dual and with vanishing curvature $R_{B}{}^{ab}=0$ by virtue of
the Maurer-Cartan equations, is

\begin{equation}
  \begin{aligned}
    \omega_{B}^{\sharp 1}
    & =
      \omega_{B}^{23}
      =
      -\tfrac{1}{2}v_{L}^{3}\,,
    \\
    & \\
    \omega_{B}^{\sharp 2}
    & =
      -\omega_{B}^{13}
      =
      -\tfrac{1}{2}v_{L}^{2}\,,
    \\
    & \\
    \omega_{B}^{\sharp 3}
    & =
      \omega_{B}^{12}
      =
      \tfrac{1}{2}v_{L}^{1}\,,    
  \end{aligned}
\end{equation}

\noindent
and the components of the tensor $\vartheta_{*}{}^{ab}$, which is also
self-dual, are

\begin{equation}
  \begin{aligned}
    \vartheta_{*}{}^{\sharp 1}
    & =
      \vartheta_{*}{}^{23}
      =
      \tfrac{1}{2}\left(f-1\right) v_{L}^{3}\,,
    \\
    & \\
    \vartheta_{*}{}^{\sharp 2}
    & =
      -\vartheta_{*}{}^{13}
      =
      -\tfrac{1}{2}\left(f^{1/2}-1\right)v_{L}^{2}\,,
    \\
    & \\
    \vartheta_{*}{}^{\sharp 3}
    & =
      \vartheta_{*}{}^{12}
      =
      \tfrac{1}{2}\left(f^{1/2}-1\right)v_{L}^{1}\,.    
  \end{aligned}
\end{equation}

The volume term in the action is 

\begin{equation}
  \begin{aligned}
    -\int_{M}  R^{ab}\wedge R_{ab}
    & =
    -24\pi^{2}\,.
  \end{aligned}
\end{equation}

Using the self-duality of $\vartheta_{*}{}^{ab}$ and the fact that $dr=0$ and,
therefore, $e^{1}=0$ on the boundaries, the GHY boundary term is

\begin{equation}
  \begin{aligned}
    -\int_{\partial M}\star (e^{a}\wedge e^{b})\wedge \vartheta_{*}{}_{ab}
    & =
      2\pi^{2}r_{0}^{2}\,.
  \end{aligned}
\end{equation}

\noindent
Notice that this result comes only from the contribution of the inner
boundary, which we are considering here and which is usually ignored. 

The other two non-vanishing boundary terms are 

\begin{equation}
  \begin{aligned}
    \int_{\partial M} R^{(+)\, ab} \wedge \vartheta_{*\, ab}
    & =
    & =
      16\pi^{2}\,.
  \end{aligned}
\end{equation}

\noindent
and

\begin{equation}
  \begin{aligned}
    \tfrac{1}{3}\int_{\partial M}\vartheta_{*}{}^{(+)\,  ab}\wedge
    \vartheta_{*\, ae}\wedge \vartheta_{*}{}^{e}{}_{b}
    & =
      8 \pi^{2}\,.
  \end{aligned}
\end{equation}

Substituting all the partial results, we get the final result

\begin{equation}
  S_{\rm ECR}^{(+)}
  =
  \frac{\pi r_{0}^{2}}{ 8G_{N}}\,,
\end{equation}

\noindent
which has no $\alpha'$ corrections. Notice that, considering only the
boundary at infinity, the vanishing action at zeroth order in
$\alpha'$ would have received non-vanishing corrections, though.

\section{Conclusions}
\label{sec-discussion}

In this work, we have studied the first-order in $\alpha'$ corrections to
gravitational instantons with self-dual curvature in the framework of the
4-dimensional Cano--Ruip\'erez Heterotic String theory effective action
\cite{Cano:2021rey}.  We solved the field equations order by order in
$\alpha'$ and found that although the metric does not receive $\alpha'$
corrections, the dilaton and axion become nontrivial due to their coupling to
the higher-order terms. We have also shown that, already to zeroth order in
$\alpha'$, it is possible to add a non-trivial combination of dilaton and axion
to these metrics without having to modify them. These solutions resemble and
generalize the D-instanton solution of Ref.~\cite{Gibbons:1995vg}.

We have obtained the generic form of the $\alpha'$ corrections for the general
Gibbons--Hawking spaces~\cite{Gibbons:1979zt,Gibbons:1987sp} and, as
particularly interesting examples, those of the Euclidean Taub--NUT and
Eguchi--Hanson solutions.

We have also devoted attention to the completion of the Euclidean
Cano--Ruip\'erez action with boundary terms so that one can formulate with it
a well-posed variational principle, and we have evaluated it for the
$\alpha'$-corrected Eguchi--Hanson instanton, finding that, if one includes the
interior boundary, the result contains no $\alpha'$ corrections whatsoever.

Interesting questions remain open. For instance, computing the action for the
Euclidean Taub--NUT metric is certainly worth exploring. This case, however,
is more subtle, as there is no clear, regular vacuum configuration to subtract
when the inner boundary is taken into account. Second, there exist interesting
gravitational instantons that are not self-dual, such as the
Taub-Bolt~\cite{Page:1978hdy} and the Chen--Teo~\cite{Chen:2011tc} ones.  The
latter is asymptotically flat, providing a natural setting in which to use the
boundary terms proposed here. Finally, the gravitational instantons found here
give an interesting starting point for testing the positive action
conjecture~\cite{Gibbons:1979xn}. We leave these and other questions for
future work.

\section*{Acknowledgments}

The work of CC is partially supported by the Agencia Nacional de Investigación y
Desarollo (ANID) through Fondecyt Regular grants No.~1240043, 1240048,
1251523, 1252053, and 1261016. The work of JLVC and TO has been supported in
part by the MCI, AEI, FEDER (UE) grant PID2024-155685NB-C21 (``Gravity,
Supergravity and Superstrings'' (GRASS)) and IFT Centro de Excelencia Severo
Ochoa CEX2020-001007-S. TO wishes to thank M.M.~Fern\'andez for her permanent
support.



\begin{thebibliography}{99}


\bibitem{Scherk:1974ca}
J.~Scherk and J.~H.~Schwarz,
``Dual Models for Nonhadrons,''
Nucl. Phys. B \textbf{81} (1974), 118-144
\doi{10.1016/0550-3213(74)90010-8}
  
\bibitem{Zwiebach:1985uq}
B.~Zwiebach,
``Curvature Squared Terms and String Theories,''
Phys. Lett. B \textbf{156} (1985), 315-317
\doi{10.1016/0370-2693(85)91616-8}


\bibitem{Callan:1985ia}
C.~G.~Callan, Jr., E.~J.~Martinec, M.~J.~Perry and D.~Friedan,
``Strings in Background Fields,''
Nucl. Phys. B \textbf{262} (1985), 593-609
doi:10.1016/0550-3213(85)90506-1
\doi{10.1016/0550-3213(85)90506-1}

\bibitem{Gross:1986iv}
D.~J.~Gross and E.~Witten,
``Superstring Modifications of Einstein's Equations,''
Nucl. Phys. B \textbf{277} (1986), 1
\doi{10.1016/0550-3213(86)90429-3}

\bibitem{Gross:1986mw}
D.~J.~Gross and J.~H.~Sloan,
``The Quartic Effective Action for the Heterotic String,''
Nucl. Phys. B \textbf{291} (1987), 41-89
\doi{10.1016/0550-3213(87)90465-2}

\bibitem{Metsaev:1987zx}
R.~R.~Metsaev and A.~A.~Tseytlin,
``Order alpha-prime (Two Loop) Equivalence of the String Equations of Motion and the Sigma Model Weyl Invariance Conditions: Dependence on the Dilaton and the Antisymmetric Tensor,''
Nucl. Phys. B \textbf{293} (1987), 385-419
\doi{10.1016/0550-3213(87)90077-0}

\bibitem{Hull:1987pc}
C.~M.~Hull and P.~K.~Townsend,
``The Two Loop Beta Function for $\sigma$ Models With Torsion,''
Phys. Lett. B \textbf{191} (1987), 115-121
\doi{10.1016/0370-2693(87)91331-1}

\bibitem{Bergshoeff:1988nn}
E.~Bergshoeff and M.~de Roo,
``Supersymmetric Chern-simons Terms in Ten-dimensions,''
Phys. Lett. B \textbf{218} (1989), 210-215
\doi{10.1016/0370-2693(89)91420-2}


\bibitem{Bergshoeff:1989de}
E.~A.~Bergshoeff and M.~de Roo,
``The Quartic Effective Action of the Heterotic String and Supersymmetry,''
Nucl. Phys. B \textbf{328} (1989), 439-468
\doi{10.1016/0550-3213(89)90336-2}

\bibitem{Cano:2021rey}
P.~A.~Cano and A.~Ruip\'erez,
``String gravity in D=4,''
Phys. Rev. D \textbf{105} (2022) no.4, 044022
\doi{10.1103/PhysRevD.105.044022}
[\arxiv{2111.04750} [hep-th]].

\bibitem{Boulware:1986dr}
D.~G.~Boulware and S.~Deser,
``Effective Gravity Theories With Dilatons,''
Phys. Lett. B \textbf{175} (1986), 409-412
\doi{10.1016/0370-2693(86)90614-3}

\bibitem{Kanti:1995vq}
P.~Kanti, N.~E.~Mavromatos, J.~Rizos, K.~Tamvakis and E.~Winstanley,
``Dilatonic black holes in higher curvature string gravity,''
Phys. Rev. D \textbf{54} (1996), 5049-5058
\doi{10.1103/PhysRevD.54.5049}
[\hepth{9511071} [hep-th]].


\bibitem{Torii:1996yi}
T.~Torii, H.~Yajima and K.~i.~Maeda,
``Dilatonic black holes with Gauss-Bonnet term,''
Phys. Rev. D \textbf{55} (1997), 739-753
\doi{10.1103/PhysRevD.55.739}
[\grqc{9606034} [gr-qc]].

\bibitem{Alexeev:1996vs}
S.~O.~Alexeev and M.~V.~Pomazanov,
``Black hole solutions with dilatonic hair in higher curvature gravity,''
Phys. Rev. D \textbf{55} (1997), 2110-2118
\doi{10.1103/PhysRevD.55.2110}
[\hepth{9605106} [hep-th]].

\bibitem{Campbell:1990fu}
B.~A.~Campbell, M.~J.~Duncan, N.~Kaloper and K.~A.~Olive,
``Gravitational dynamics with Lorentz Chern-Simons terms,''
Nucl. Phys. B \textbf{351} (1991), 778-792
\doi{10.1016/S0550-3213(05)80045-8}

\bibitem{Jackiw:2003pm}
R.~Jackiw and S.~Y.~Pi,
``Chern-Simons modification of general relativity,''
Phys. Rev. D \textbf{68} (2003), 104012
\doi{10.1103/PhysRevD.68.104012}
[\grqc{0308071} [gr-qc]].

\bibitem{Alexander:2009tp}
S.~Alexander and N.~Yunes,
``Chern-Simons Modified General Relativity,''
Phys. Rept. \textbf{480} (2009), 1-55
\doi{10.1016/j.physrep.2009.07.002}
[\arxiv{0907.2562} [hep-th]].

\bibitem{Grumiller:2007rv}
D.~Grumiller and N.~Yunes,
``How do Black Holes Spin in Chern-Simons Modified Gravity?,''
Phys. Rev. D \textbf{77} (2008), 044015
\doi{10.1103/PhysRevD.77.044015}
[\arxiv{0711.1868} [gr-qc]].

\bibitem{Konno:2009kg}
K.~Konno, T.~Matsuyama and S.~Tanda,
``Rotating black hole in extended Chern-Simons modified gravity,''
Prog. Theor. Phys. \textbf{122} (2009), 561-568
\doi{10.1143/PTP.122.561}
[\arxiv{0902.4767} [gr-qc]].

\bibitem{Ahmedov:2010fz}
H.~Ahmedov and A.~N.~Aliev,
``Black String and Godel type Solutions of Chern-Simons Modified Gravity,''
Phys. Rev. D \textbf{82} (2010), 024043
\doi{10.1103/PhysRevD.82.024043}
[\arxiv{1003.6017} [hep-th]].

\bibitem{Brihaye:2016lsx}
Y.~Brihaye and E.~Radu,
``Remarks on the Taub--NUT solution in Chern-Simons modified gravity,''
Phys. Lett. B \textbf{764} (2017), 300-305
\doi{10.1016/j.physletb.2016.11.055}
[\arxiv{1610.09952} [gr-qc]].

\bibitem{Cisterna:2018jsx}
A.~Cisterna, C.~Corral and S.~del Pino,
``Static and rotating black strings in dynamical Chern{\textendash}Simons modified gravity,''
Eur. Phys. J. C \textbf{79} (2019) no.5, 400
\doi{10.1140/epjc/s10052-019-6910-5}
[\arxiv{1809.02903} [gr-qc]].

\bibitem{Nashed:2023qjm}
G.~G.~L.~Nashed and S.~Nojiri,
``Slow-rotating charged black hole solution in
dynamical Chern-Simons modified gravity,''
Phys. Rev. D \textbf{107} (2023) no.6, 064069
\doi{10.1103/PhysRevD.107.064069}
[\arxiv{2303.07349} [gr-qc]].

\bibitem{Anabalon:2024abz}
  A.~Anabal\'on, D.~Astefanesei, A.~Cisterna, F.~Izaurieta,
  J.~Oliva, C.~Quijada and C.~Quinzacara,
``Rotating and accelerating AdS black holes in Einstein-Gauss-Bonnet gravity,''
Phys. Lett. B \textbf{857} (2024), 139000
\doi{10.1016/j.physletb.2024.139000}
[\arxiv{2404.04691} [hep-th]].

\bibitem{Tapia:2025vtn}
  L.~Tapia, M.~Aguayo, A.~Anabal\'on, D.~Astefanesei, N.~Grandi,
  F.~Izaurieta, J.~Oliva and C.~Quinzacara,
``(Quasi-)normal modes of rotating black holes and
new solitons in Einstein-Gauss-Bonnet,''
Phys. Lett. B \textbf{862} (2025), 139347
\doi{10.1016/j.physletb.2025.139347}
[\arxiv{2411.08001} [hep-th]].

\bibitem{Dehghani:2002wn}
M.~H.~Dehghani,
``Charged rotating black branes in anti-de Sitter Einstein-Gauss-Bonnet gravity,''
Phys. Rev. D \textbf{67} (2003), 064017
\doi{10.1103/PhysRevD.67.064017}
[\hepth{0211191} [hep-th]].

\bibitem{Dehghani:2003ea}
M.~H.~Dehghani,
``Magnetic branes in Gauss-Bonnet gravity,''
Phys. Rev. D \textbf{69} (2004), 064024
\doi{10.1103/PhysRevD.69.064024}
[\hepth{0312030} [hep-th]].

\bibitem{Dehghani:2006cu}
M.~H.~Dehghani, G.~H.~Bordbar and M.~Shamirzaie,
``Thermodynamics of Rotating Solutions in Gauss-Bonnet-Maxwell
Gravity and the Counterterm Method,''
Phys. Rev. D \textbf{74} (2006), 064023
\doi{10.1103/PhysRevD.74.064023}
[\hepth{0607067} [hep-th]].

\bibitem{Hendi:2010zza}
S.~H.~Hendi and B.~E.~Panah,
``Thermodynamics of rotating black branes in
Gauss-Bonnet-nonlinear Maxwell gravity,''
Phys. Lett. B \textbf{684} (2010), 77-84
\doi{10.1016/j.physletb.2010.01.026}
[\arxiv{1008.0102} [hep-th]].

\bibitem{Cano:2023dyg}
P.~A.~Cano and M.~David,
``The extremal Kerr entropy in higher-derivative gravities,''
JHEP \textbf{05} (2023), 219
\doi{10.1007/JHEP05(2023)219}
[\arxiv{2303.13286} [hep-th]].

\bibitem{Ortin:2024emt}
T.~Ort\'{\i}n and M.~Zatti,
``On the thermodynamics of the black holes of the
Cano--Ruip\'erez 4-dimensional string effective action,''
JHEP \textbf{06} (2025), 026
\doi{10.1007/JHEP06(2025)026}
[\arxiv{2411.10417} [hep-th]].

\bibitem{Ortin:2021ade}
T.~Ort\'{\i}n,
``Komar integrals for theories of higher order
in the Riemann curvature and black-hole chemistry,''
JHEP \textbf{08} (2021), 023
\doi{10.1007/JHEP08(2021)023}

\bibitem{Zatti:2023oiq}
M.~Zatti,
``{\ensuremath{\alpha}}$^{\prime}$ corrections to
4-dimensional non-extremal stringy black holes,''
JHEP \textbf{11} (2023), 185
\doi{10.1007/JHEP11(2023)185}
[\arxiv{2308.12879} [hep-th]].


\bibitem{Meessen:2022hcg}
P.~Meessen, D.~Mitsios and T.~Ort\'{\i}n,
``Black hole chemistry, the cosmological constant and the embedding tensor,''
JHEP \textbf{12} (2022), 155
\doi{10.1007/JHEP12(2022)155}
[\arxiv{2203.13588} [hep-th]].

\bibitem{Agurto-Sepulveda:2022vvf}
F.~Agurto-Sep{\'u}lveda, M.~Chernicoff, G.~Giribet, J.~Oliva and M.~Oyarzo,
``Slowly rotating and accelerating {\ensuremath{\alpha}}'-corrected black holes in four and higher dimensions,''
Phys. Rev. D \textbf{107} (2023) no.8, 084014
\doi{10.1103/PhysRevD.107.084014}
[\arxiv{2207.13214} [hep-th]].

\bibitem{Agurto-Sepulveda:2024iwu}
F.~Agurto-Sep{\'u}lveda, J.~Oliva, M.~Oyarzo and D.~R.~G.~Schleicher,
``Black hole shadows of {\ensuremath{\alpha}}'-corrected black holes,''
Phys. Rev. D \textbf{110} (2024) no.2, 024078
\doi{10.1103/PhysRevD.110.024078}
[\arxiv{2404.12811} [gr-qc]].

\bibitem{Cano:2018qev}
P.~A.~Cano, P.~Meessen, T.~Ort\'{\i}n and P.~F.~Ram\'{\i}rez,
``$\alpha'$-corrected black holes in String Theory,''
JHEP \textbf{05} (2018), 110
\doi{10.1007/JHEP05(2018)110}
[arXiv:1803.01919 [hep-th]].

\bibitem{Chimento:2018kop}
S.~Chimento, P.~Meessen, T.~Ort\'{\i}n, P.~F.~Ram\'{\i}rez and A.~Ruip\'erez,
``On a family of $\alpha'$-corrected solutions
of the Heterotic Superstring effective action,''
JHEP \textbf{07} (2018), 080
\doi{10.1007/JHEP07(2018)080}
[\arxiv{1803.04463} [hep-th]].

\bibitem{Cano:2018brq}
P.~A.~Cano, S.~Chimento, P.~Meessen, T.~Ort\'{\i}n, P.~F.~Ram\'{\i}rez and A.~Ruip\'erez,
``Beyond the near-horizon limit: Stringy corrections to Heterotic Black Holes,''
JHEP \textbf{02} (2019), 192
\doi{10.1007/JHEP02(2019)192}
[\arxiv{1808.03651} [hep-th]].

\bibitem{Cano:2019ycn}
P.~A.~Cano, S.~Chimento, R.~Linares, T.~Ort\'{\i}n and P.~F.~Ram\'{\i}rez,
``$\alpha'$ corrections of Reissner-Nordstr{\"o}m black holes,''
JHEP \textbf{02} (2020), 031
\doi{10.1007/JHEP02(2020)031}
[\arxiv{1910.14324} [hep-th]].

\bibitem{Cano:2021nzo}
P.~A.~Cano, T.~Ort\'{\i}n, A.~Ruip\'erez and M.~Zatti,
``Non-supersymmetric black holes with {\ensuremath{\alpha}}' corrections,''
JHEP \textbf{03} (2022), 103
\doi{10.1007/JHEP03(2022)103}
[\arxiv{2111.15579} [hep-th]].

\bibitem{Ortin:2021win}
T.~Ort\'{\i}n, A.~Ruip\'erez and M.~Zatti,
``Extremal stringy black holes in equilibrium at first order in $\alpha'$,''
SciPost Phys. Core \textbf{6} (2023) no.4, 072
\doi{10.21468/SciPostPhysCore.6.4.072}
[\arxiv{2112.12764} [hep-th]].

\bibitem{Cano:2022tmn}
P.~A.~Cano, T.~Ort\'{\i}n, A.~Ruip\'erez and M.~Zatti,
``Non-extremal, {\ensuremath{\alpha}}'-corrected black holes
in 5-dimensional heterotic superstring theory,''
JHEP \textbf{12} (2022), 150
\doi{10.1007/JHEP12(2022)150}
[\arxiv{2210.01861} [hep-th]].

\bibitem{Hu:2025aji}
P.~J.~Hu, L.~Ma, Y.~Pang and R.~J.~Saskowski,
``Higher-derivative heterotic Kerr-Sen black holes,''
JHEP \textbf{02} (2026), 235
\doi{10.1007/JHEP02(2026)235}
[\arxiv{2506.20077} [hep-th]].

\bibitem{Dehghani:2005zm}
M.~H.~Dehghani and R.~B.~Mann,
``NUT-charged black holes in Gauss-Bonnet gravity,''
Phys. Rev. D \textbf{72} (2005), 124006
\doi{10.1103/PhysRevD.72.124006}
[\hepth{0510083} [hep-th]].

\bibitem{Dehghani:2006dh}
M.~H.~Dehghani and R.~B.~Mann,
``Thermodynamics of rotating charged black branes
in third order Lovelock gravity and the counterterm method,''
Phys. Rev. D \textbf{73} (2006), 104003
\doi{10.1103/PhysRevD.73.104003}
[\hepth{0602243} [hep-th]].

\bibitem{Dehghani:2006aa}
M.~H.~Dehghani and S.~H.~Hendi,
``Taub--NUT/bolt black holes in Gauss-Bonnet-Maxwell gravity,''
Phys. Rev. D \textbf{73} (2006), 084021
\doi{10.1103/PhysRevD.73.084021}
[\hepth{0602069} [hep-th]].

\bibitem{Hendi:2008wq}
S.~H.~Hendi and M.~H.~Dehghani,
``Taub--NUT Black Holes in Third order Lovelock Gravity,''
Phys. Lett. B \textbf{666} (2008), 116-120
\doi{10.1016/j.physletb.2008.07.002}
[\arxiv{0802.1813} [hep-th]].


\bibitem{Corral:2019leh}
C.~Corral, D.~Flores-Alfonso and H.~Quevedo,
``Charged Taub--NUT solution in Lovelock gravity
with generalized Wheeler polynomials,''
Phys. Rev. D \textbf{100} (2019) no.6, 064051
\doi{10.1103/PhysRevD.100.064051}
[\arxiv{1908.06908} [gr-qc]].


\bibitem{Corral:2022udb}
C.~Corral, D.~Flores-Alfonso, G.~Giribet and J.~Oliva,
``Higher-curvature generalization of Eguchi--Hanson spaces,''
Phys. Rev. D \textbf{106} (2022) no.8, 084055
\doi{10.1103/PhysRevD.106.084055}
[\arxiv{2207.04014} [hep-th]].

\bibitem{Corral:2025yvr}
C.~Corral, B.~Diez, D.~Flores-Alfonso, N.~Merino and L.~Sanhueza,
``Inhomogeneous metrics on complex bundles in Lovelock gravity,''
Phys. Rev. D \textbf{111} (2025) no.12, 124016
\doi{10.1103/PhysRevD.111.124016}
[\arxiv{2504.11562} [hep-th]].


\bibitem{Gibbons:1979zt}
G.~W.~Gibbons and S.~W.~Hawking,
``Gravitational Multi - Instantons,''
Phys.\ Lett.\ B {\bf 78} (1978) 430.
\doi{10.1016/0370-2693(78)90478-1}

\bibitem{Gibbons:1987sp}
G.~W.~Gibbons and P.~J.~Ruback,
``The Hidden Symmetries of Multicenter Metrics,''
Commun.\ Math.\ Phys.\  {\bf 115} (1988) 267.
\doi{10.1007/BF01466773}

\bibitem{tHooft:1976rip}
G.~'t Hooft,
``Symmetry Breaking Through Bell-Jackiw Anomalies,''
Phys. Rev. Lett. \textbf{37} (1976), 8-11
\doi{10.1103/PhysRevLett.37.8}

\bibitem{tHooft:1976snw}
G.~'t Hooft,
``Computation of the Quantum Effects Due to a Four-Dimensional Pseudoparticle,''
Phys. Rev. D \textbf{14} (1976), 3432-3450
[erratum: Phys. Rev. D \textbf{18} (1978), 2199]
\doi{10.1103/PhysRevD.14.3432}

\bibitem{Taub:1950ez}
A.~H.~Taub,
``Empty space-times admitting a three-parameter group of motions,''
Annals Math. \textbf{53} (1951), 472-490
\doi{10.2307/1969567}

\bibitem{Newman:1963yy}
E.~Newman, L.~Tamburino and T.~Unti,
``Empty space generalization of the Schwarzschild metric,''
J. Math. Phys. \textbf{4} (1963), 915
\doi{10.1063/1.1704018}

\bibitem{Hawking:1976jb}
S.~W.~Hawking,
``Gravitational Instantons,''
Phys. Lett. A \textbf{60} (1977), 81
\doi{10.1016/0375-9601(77)90386-3}

\bibitem{Eguchi:1977iu}
T.~Eguchi, P.~B.~Gilkey and A.~J.~Hanson,
``Is the Taub--NUT Metric a Gravitational Instanton?,''
Phys. Rev. D \textbf{17} (1978), 423-427
\doi{10.1103/PhysRevD.17.423}

\bibitem{Eguchi:1978xp}
T.~Eguchi and A.~J.~Hanson,
``Asymptotically Flat Selfdual Solutions to Euclidean Gravity,''
Phys. Lett. B \textbf{74} (1978), 249-251
\doi{10.1016/0370-2693(78)90566-X}

\bibitem{Eguchi:1978gw}
T.~Eguchi and A.~J.~Hanson,
``Selfdual Solutions to Euclidean Gravity,''
Annals Phys. \textbf{120} (1979), 82
\doi{10.1016/0003-4916(79)90282-3}

\bibitem{Eguchi:1980jx}
T.~Eguchi, P.~B.~Gilkey and A.~J.~Hanson,
``Gravitation, Gauge Theories and Differential Geometry,''
Phys. Rept. \textbf{66} (1980), 213
\doi{10.1016/0370-1573(80)90130-1}

\bibitem{Gibbons:1995vg}
G.~W.~Gibbons, M.~B.~Green and M.~J.~Perry,
``Instantons and seven-branes in type IIB superstring theory,''
Phys. Lett. B \textbf{370} (1996), 37-44
\doi{10.1016/0370-2693(95)01565-5}
[\hepth{9511080} [hep-th]].

\bibitem{Ortin:2015hya}
T.~Ort\'{\i}n,
``Gravity and Strings'', 2nd edition, 
Cambridge University Press, 2015.

\bibitem{Gomez-Fayren:2023qly}
C.~G\'omez-Fayr\'en, P.~Meessen and T.~Ort\'{\i}n,
``Covariant generalized conserved charges of General Relativity,''
JHEP \textbf{09} (2023), 174
\doi{10.1007/JHEP09(2023)174}
[\arxiv{2307.04041} [gr-qc]].

\bibitem{Aros:1999id}
R.~Aros, M.~Contreras, R.~Olea, R.~Troncoso and J.~Zanelli,
``Conserved charges for gravity with locally AdS asymptotics,''
Phys. Rev. Lett. \textbf{84} (2000), 1647-1650
\doi{10.1103/PhysRevLett.84.1647}
[\grqc{9909015} [gr-qc]].

\bibitem{Olea:2005gb}
R.~Olea,
``Mass, angular momentum and thermodynamics in four-dimensional Kerr-AdS black holes,''
JHEP \textbf{06} (2005), 023
\doi{10.1088/1126-6708/2005/06/023}
[\hepth{0504233} [hep-th]].

\bibitem{Anastasiou:2020zwc}
G.~Anastasiou, O.~Miskovic, R.~Olea and I.~Papadimitriou,
``Counterterms, Kounterterms, and the variational problem in AdS gravity,''
JHEP \textbf{08} (2020), 061
\doi{10.1007/JHEP08(2020)061}
[\arxiv{2003.06425} [hep-th]].

\bibitem{Miskovic:2009bm}
O.~Miskovic and R.~Olea,
``Topological regularization and self-duality in four-dimensional anti-de Sitter gravity,''
Phys. Rev. D \textbf{79} (2009), 124020
\doi{10.1103/PhysRevD.79.124020}
[\arxiv{0902.2082} [hep-th]].

\bibitem{Araneda:2016iiy}
R.~Araneda, R.~Aros, O.~Miskovic and R.~Olea,
``Magnetic Mass in 4D AdS Gravity,''
Phys. Rev. D \textbf{93} (2016) no.8, 084022
\doi{10.1103/PhysRevD.93.084022}
[\arxiv{1602.07975} [hep-th]].

\bibitem{Corral:2024lva}
C.~Corral and R.~Olea,
``Electric-magnetic duality of dyonic Kerr-Newman-NUT-AdS spacetimes,''
Phys. Rev. D \textbf{110} (2024) no.10, 104021
\doi{10.1103/PhysRevD.110.104021}
[\arxiv{2408.03901} [hep-th]].

\bibitem{Gross:1983hb}
D.~J.~Gross and M.~J.~Perry,
``Magnetic Monopoles in Kaluza-Klein Theories,''
Nucl. Phys. B \textbf{226} (1983), 29-48
\doi{10.1016/0550-3213(83)90462-5}

\bibitem{Sorkin:1983ns}
R.~D.~Sorkin,
``Kaluza-Klein Monopole,''
Phys. Rev. Lett. \textbf{51} (1983), 87-90
\doi{10.1103/PhysRevLett.51.87}

\bibitem{Prasad:1979kg}
M.~K.~Prasad,
``Equivalence of Eguchi--Hanson metric to two-center Gibbons--Hawking metric,''
Phys. Lett. B \textbf{83} (1979), 310-310
\doi{10.1016/0370-2693(79)91114-6}

\bibitem{Pedersen:1985}
H.~Pedersen,
``Eguchi--Hanson metrics with cosmological constant,''
Class. and Quan. Gravity \textbf{2} (1985), 579
\doi{10.1088/0264-9381/2/4/022}

\bibitem{Clarkson:2005qx}
R.~Clarkson and R.~B.~Mann,
``Eguchi--Hanson solitons in odd dimensions,''
Class. Quant. Grav. \textbf{23} (2006), 1507-1524
\doi{10.1088/0264-9381/23/5/005}
[\hepth{0508200} [hep-th]].

\bibitem{Clarkson:2006zk}
R.~Clarkson and R.~B.~Mann,
``Soliton solutions to the Einstein equations in five dimensions,''
Phys. Rev. Lett. \textbf{96} (2006), 051104
\doi{10.1103/PhysRevLett.96.051104}
[\hepth{0508109} [hep-th]].

\bibitem{Chng:2006gh}
B.~Chng, R.~B.~Mann and C.~Stelea,
``Accelerating Taub--NUT and Eguchi--Hanson solitons in four dimensions,''
Phys. Rev. D \textbf{74} (2006), 084031
\doi{10.1103/PhysRevD.74.084031}
[\grqc{0608092} [gr-qc]].

\bibitem{Wong:2011aa}
A.~W.~C.~Wong and R.~B.~Mann,
``Five-Dimensional Eguchi--Hanson Solitons in Einstein-Gauss-Bonnet Gravity,''
Phys. Rev. D \textbf{85} (2012), 046010
\doi{10.1103/PhysRevD.85.046010}
[\arxiv{1112.2229} [hep-th]].

\bibitem{Durgut:2022xzw}
T.~Durgut, R.~A.~Hennigar, H.~K.~Kunduri and R.~B.~Mann,
``Phase transitions and stability of Eguchi--Hanson-AdS solitons,''
JHEP \textbf{03} (2023), 114
\doi{10.1007/JHEP03(2023)114}
[\arxiv{2212.12685} [gr-qc]].

\bibitem{York:1972sj}
J.~W.~York, Jr.,
``Role of conformal three geometry in the dynamics of gravitation,''
Phys. Rev. Lett. \textbf{28} (1972), 1082-1085
\doi{10.1103/PhysRevLett.28.1082}

\bibitem{Gibbons:1976ue}
G.~W.~Gibbons and S.~W.~Hawking,
``Action Integrals and Partition Functions in Quantum Gravity,''
Phys. Rev. D \textbf{15} (1977), 2752-2756
\doi{10.1103/PhysRevD.15.2752}

\bibitem{Page:1978hdy}
D.~N.~Page,
``Taub - Nut Instanton With an Horizon,''
Phys. Lett. B \textbf{78} (1978), 249-251
\doi{10.1016/0370-2693(78)90016-3}


\bibitem{Chen:2011tc}
Y.~Chen and E.~Teo,
``A New AF gravitational instanton,''
Phys. Lett. B \textbf{703} (2011), 359-362
\doi{10.1016/j.physletb.2011.07.076}
[\arxiv{1107.0763} [gr-qc]].

\bibitem{Gibbons:1979xn}
G.~W.~Gibbons and C.~N.~Pope,
``The Positive Action Conjecture and Asymptotically
Euclidean Metrics in Quantum Gravity,''
Commun. Math. Phys. \textbf{66} (1979), 267-290
\doi{10.1007/BF01197188}






\end{thebibliography}
\end{document}